\PassOptionsToPackage{prologue,svgnames,table}{xcolor}
\documentclass[sigplan,twocolumn,10pt]{acmart}

\newif\ifARXIV
\newif\ifSUB

\usepackage{hyperref}

\usepackage{amsmath,amsfonts}
\usepackage{algorithm}
\usepackage{algpseudocode}
\usepackage{graphicx}
\usepackage{textcomp}
\usepackage{regexpatch}
\usepackage{enumitem}
\usepackage{subcaption}
\usepackage{multirow}
\usepackage{xspace}
\usepackage{orcidlink}
\usepackage{doi}
\usepackage{xspace}
\usepackage{xargs}
\usepackage{microtype}
\usepackage{tabularx}
\usepackage{booktabs}
\usepackage{listings}
\usepackage[frozencache,cachedir=minted-cache]{minted} 
\usepackage{parcolumns}
\usepackage[many]{tcolorbox}
\usepackage{mathtools}
\usepackage[export]{adjustbox}
\usepackage{float}
\usepackage{makecell}
\usepackage{wrapfig}

\usepackage[absolute]{textpos}

\newcommand{\mcrot}[4]{\multicolumn{#1}{#2}{\rlap{\rotatebox{#3}{#4}~}}}

\makeatletter
\begingroup\endlinechar=-1\relax
       \everyeof{\noexpand}%
       \edef\x{\endgroup\def\noexpand\homepath{%
         \@@input|"kpsewhich --var-value=HOME" }}\x
\makeatother

\usepackage{csquotes}

\newcommand{\toolname}{SeBS-Flow}

\newcommand{\changed}[1]{\textcolor{black}{#1}}
\newcommand{\eurosys}[1]{\textcolor{black}{#1}}
\newcommand{\cameraready}[1]{\textcolor{black}{#1}}

\newcommand{\caseicon}[1]{
  \texorpdfstring{
    \protect\raisebox{
      -.2\baselineskip
    }{
    \protect\hspace{-0.1cm}\includegraphics[scale=0.39,trim=0.1 0.1 0.1 0.1, clip]{#1}\hspace{-0.05cm}
    }
  }{}
}

\setcopyright{none}
\renewcommand\footnotetextcopyrightpermission[1]{}

\copyrightyear{2025}
\acmYear{2025}
\setcopyright{cc}
\setcctype{by}
\acmConference[EuroSys '25]{Twentieth European Conference on Computer Systems}{March 30-April 3, 2025}{Rotterdam, Netherlands}
\acmBooktitle{Twentieth European Conference on Computer Systems (EuroSys '25), March 30-April 3, 2025, Rotterdam, Netherlands}\acmDOI{10.1145/3689031.3717465}
\acmISBN{979-8-4007-1196-1/2025/03}

\begin{document}
\ARXIVtrue
\SUBfalse

\begin{textblock}{14}(1,0.5)

\noindent \begin{center}\textbf{This is the author's version of the paper submitted and accepted at EuroSys'25.\\The definitive Version of Record is published here: \url{https://doi.org/10.1145/3689031.3717465}.} 
\end{center}
\end{textblock}

\title{\toolname: Benchmarking Serverless Cloud Function Workflows}

\author{Larissa Schmid}
 \orcid{0000-0002-3600-6899}
 \affiliation{%
   \institution{Karlsruhe Institute of Technology}
   \country{Germany}
}

\author{Marcin Copik}
\orcid{0000-0002-7606-5519}
\affiliation{%
  \institution{ETH Zurich}
  \country{Switzerland}
  }
  
\author{Alexandru Calotoiu}
\orcid{0000-0001-9095-9108}
\affiliation{%
  \institution{ETH Zurich}
  \country{Switzerland}
  }

\author{Laurin Brandner}
\orcid{0009-0005-8251-9117}
\affiliation{%
  \institution{ETH Zurich}
  \country{Switzerland}
  }

\author{Anne Koziolek}
\orcid{0000-0002-1593-3394}
\affiliation{%
  \institution{Karlsruhe Institute of Technology}
  \country{Germany}
}

\author{Torsten Hoefler}
\orcid{0000-0002-1333-9797}
\affiliation{%
  \institution{ETH Zurich}
  \country{Switzerland}
  }

\renewcommand{\shortauthors}{Schmid et al.}

\begin{abstract}
Serverless computing has emerged as a prominent paradigm, with a significant adoption rate among cloud customers.
While this model offers advantages such as abstraction from the deployment and resource scheduling, it also poses limitations in handling complex use cases due to the restricted nature of individual functions.
Serverless workflows address this limitation by orchestrating multiple functions into a cohesive application.
However, existing serverless workflow platforms exhibit significant differences in their programming models and infrastructure, making fair and consistent performance evaluations difficult in practice. 
To address this gap, we propose the first serverless workflow benchmarking suite \toolname{}, providing a platform-agnostic workflow model that enables consistent benchmarking across various platforms. \toolname{} includes six real-world application benchmarks and four microbenchmarks representing different computational patterns. 
We conduct comprehensive evaluations on three major cloud platforms, assessing performance, cost, scalability, and runtime deviations. We make our benchmark suite open-source, enabling rigorous and comparable evaluations of serverless workflows over time. 

\noindent \small{\textbf{Implementation}: \url{https://github.com/spcl/serverless-benchmarks}}

\noindent \small{\textbf{Artifact}: \url{https://github.com/spcl/sebs-flow-artifact}}
\end{abstract}

\ccsdesc[500]{Computer systems organization~Cloud computing}
\ccsdesc[500]{Networks~Cloud computing}
\ccsdesc[500]{Software and its engineering~Software performance}
\ccsdesc[300]{General and reference~Performance}
\ccsdesc[300]{General and reference~Metrics}
\ccsdesc[300]{General and reference~Measurement}
\ccsdesc[300]{General and reference~Evaluation}

\keywords{benchmark, serverless, function-as-a-service, faas, workflow, orchestration, serverless DAG}

\maketitle

\section{Introduction}

Serverless computing gained major adoption in the industry~\cite{eismann2021-why,LEITNER2019340}, with 50-70\% of cloud customers using serverless functions and containers~\cite{stateOfServerless}.
In the Function-as-a-Service (FaaS) programming model, developers implement stateless functions and invoke them through a REST interface.
The actual function deployment and resource scheduling becomes the responsibility of the cloud operator: 
Developers are no longer concerned with managing their applications 
and are charged only for resources used to handle function invocations.
While the primitiveness of FaaS can be an important benefit~\cite{eismann2021-why}, it is also a major drawback: a single function is insufficient to cover all use cases.
Functions must be composed to build larger applications, keep the design modular, 
or use pre-defined and standardized functions, e.g., for machine learning inference.

Serverless workflows allow to 
chain and aggregate multiple functions into a single application by creating a graph of functions and automating the execution of a sequence through control and data dependencies.
They include control-flow components - conditions and loops - which allows them to represent full computations such as multi-stage machine learning pipelines. 
Developers implement functions and define the workflow structure in a cloud-specific format.
Cloud operators then control the workflow invocation and orchestration, retaining the ability to optimize resource consumption, e.g., through optimized function placement, oversubscription, targeting idle resources, and co-locating functions that depend on each other~\cite{mahgoub2021sonic,10.5555/3277355.3277444,copik2022softwarepaper}.

Workflows have been adopted by the most popular commercial cloud platforms~\cite{awsstep,azuredurable,gcpworkflows} and make up almost a third of serverless applications~\cite{eismann2021review}. 
However, just like every FaaS platform is different~\cite{copik2021sebs}, serverless workflows are quite distinct from each other. 
Not only the different APIs and incompatible graph syntax and format complicate the software development process, but also fundamentally different programming models: workflow platforms diverge in the statelessness of functions and the static nature of graph definition (Section~\ref{sec:platforms}).
Even though FaaS platforms might seem like the same product, they offer drastically different performance, reliability, and cost~\cite{copik2021sebs,10.1145/3401025.3401738,10.5555/3277355.3277369}.
With workflows built as an orchestration of 
functions, their functionality and performance is affected by both orchestration service and existing differences in the underlying compute infrastructure.
As such details are hidden, an information gap between developers and providers arises~\cite{wen2023}.
Thus, the software developers need to conduct extensive performance testing
of the cloud services to estimate the performance of their workloads and understand platform limitations up-front, as choosing a certain platform implies significant lock-in~\cite{adzic2017}, with only limited support for testing~\cite{LEITNER2019340}. 

\begin{table}[t]
\begin{adjustbox}{width=.95\linewidth}
\begin{tabular}{l|c|cccccc|ccccc|c}
& &\multicolumn{6}{l}{Benchmarks} & \multicolumn{5}{c}{Platforms} & \\
Papers & Total & \mcrot{1}{l}{60}{Micro} & \mcrot{1}{l}{60}{Webapp} & \mcrot{1}{l}{60}{Multimedia} & \mcrot{1}{l}{60}{Data Proc.} & \mcrot{1}{l}{60}{ML} & \mcrot{1}{l}{60}{Scientific} & \mcrot{1}{l}{60}{AWS} & \mcrot{1}{l}{60}{Azure} & \mcrot{1}{l}{60}{GCP} & \mcrot{1}{l}{60}{Other} & \mcrot{1}{l}{60}{Research} & \mcrot{1}{l}{60}{Artifact?}\\
\hline
Analysis & 14 & 7 & 1 & 4 & 2 & 4 & 2 & 8 & 4 & 3 & 3 & 3 & 5 \\
Optimization & 17 & 8 & 3 & 4 & 4 & 5 & 6 & 9 & 0 & 2 & 2 & 7 & 4 \\
Application & 18 & 1 & 4 & 1 & 4 & 1 & 7 & 15 & 5 & 5 & 2 & 3 & 9\\
Prog. Model & 23 & 10 & 6 & 5 & 8 & 11 & 8 & 10 & 3 & 1 & 2 & 16 & 11\\
\end{tabular}
\end{adjustbox}
\caption{Analysis of 72 research papers on serverless workflows with benchmarks.}
\label{tab:table_intro}
\end{table}

We propose the first \textbf{serverless workflows benchmarking suite} to support software developers and the quickly growing research activity in serverless workflows.
Our work provides a baseline and benchmarking methodology for evaluating and comparing the performance of workflows on different platforms, highlighting their strengths and weaknesses.
We examined 72 different research contributions to determine the similarity of their evaluation baselines (Table~\ref{tab:table_intro}).
We found that publications use different applications to benchmark the performance of new ideas, do not cover the same classes of workloads, and do not always compare against the same subset of platforms.
Without a consistent baseline, comparing research results and establishing the most promising ideas becomes impossible~\cite{SCHEUNER2020110708}.
Benchmarking suites and systems have been proposed for FaaS~\cite{copik2021sebs,10.1145/3401025.3401738,9027346,Kim2019}, but 
a benchmarking suite for serverless workflows has remained an open problem.
A comprehensive, consistent, platform-independent, and portable benchmarking suite will support the ongoing research work~\cite{papadopoulos2021,SCHEUNER2020110708} and enable developers to differentiate between alternative solutions. 
We establish a unified and portable \textbf{workflow model} to abstract away the differences between different platforms (Section~\ref{sec:model}). 
We design the \textbf{benchmarking suite} (Section~\ref{sec:impl}) and include \textbf{six workflow benchmarks} based on solutions common in research and industry (Section~\ref{sec:benchmarks}). 
Applications are implemented in our unified workflow model, providing an identical benchmark structure for each platform. 
We evaluate expressiveness and overhead of our model (Section~\ref{sec:evaluation-model})
and use our benchmarking suite to comprehensively evaluate the three major cloud workflow services (Section~\ref{sec:evaluation}). 
We follow the FAIR principle~\cite{wilkinson2016fair} and release our benchmark suite on an open-source license, enabling automatic repetition of our experiments, allowing reproducible results, and measuring performance changes in clouds over time. 
We make the following contributions:
\begin{itemize}[leftmargin=*]
    \item We introduce a platform-agnostic workflow definition, 
    automatically transcribe the application into a cloud's proprietary presentations, and enable developers to run near identical workloads on different systems.
    \item We propose a benchmark suite with six real-world application benchmarks and four microbenchmarks.
    \item We extensively analyze performance, cost, scaling, and stability of three major cloud platforms. 
\end{itemize}
\section{Background}

Serverless workflows introduce multiple new challenges to the software development process due to differences in the workflows platforms (Section~\ref{sec:platforms}).
To model workflows, we use the formalism and semantics of Petri Nets (Section~\ref{sec:background:workflow-nets}).

\subsection{Developing Serverless Workflows}
\label{sec:platforms}

\begin{table}[t]
    \centering
    \begin{adjustbox}{width=\linewidth}
    \begin{tabular}{ll!{\vrule width -5pt}lll}
    \specialrule{.1em}{0em}{0em} 
        Platform & Prog. Model & Model Flexibility & Max. Parallelism& Interface \\\hline
        AWS & State Machine & Static & 40 & JSON \\
        \rowcolor{Gainsboro!50} 
        Azure & \makecell[l]{Orchestrator \\ Function} & Dynamic & Unlimited & \makecell[l]{Durable \\ Functions} \\ 
        Google & State Machine & Semi-dynamic & 20 & JSON/YAML \\
    \specialrule{.1em}{0em}{0em} 
    \end{tabular}
    \end{adjustbox}
    \caption{Key features of serverless workflows platforms.}
    \label{tab:platform-features}
\end{table}

While software engineers are increasingly interested in serverless applications~\cite{wen2021}, they encounter a wide range of challenges while developing them, with the first questions about the different capabilities of the platforms arising before starting the implementation~\cite{wen2021,sampe2021}: 
Workflows have been adopted by all major cloud providers, but their implementations are significantly different in capabilities 
(Table~\ref{tab:platform-features}).
We focus on AWS Step Functions, Google Cloud Workflows, and Azure Durable Functions, as they play a leading role. 

\begin{figure}
    \begin{subfigure}[b]{\linewidth}
      \scriptsize
\begin{minted}{python}
tasks = []
for i in range(4):
  tasks.append(context.call_activity("process", i)
res = yield context.task_all(parallel_tasks)
\end{minted}
      \caption{Azure Durable Functions}
      \label{lst:loop-azure}
    \end{subfigure}\\
    \begin{subfigure}[b]{0.49\linewidth}
      \scriptsize
\begin{minted}{json}
"assign_array": {
    "assign": [
      {"array": [0, 1, 2, 3]}]
},
"process": {
    "call":"exp.exec.map",
    "args":{
      "workflow_id":"map",
      "arguments":"${array}"
    },
    "result":"res" 
}
"separate map-workflow:"
"main": {
"params":[ "elem" ],
"steps":[
  {
  "map":{
    "call":"http.post",
    "args":{
      "url":"google.process",
      "body":{
        "payload":"${elem}"
      } 
    },
    "result":"elem" } 
  },
  { "ret": { 
      "return":
      "${elem.body}" } }
] } 
\end{minted}
      \caption{Google Cloud Workflows}
      \label{lst:loop-google}
    \end{subfigure}
    \hfill
    \begin{subfigure}[b]{0.49\linewidth}
      \scriptsize
\begin{minted}{json}
"init": {
  "Type": "Pass",
  "Result": "States.Array(0, 
            1, 2, 3)",
  "ResultPath": "$.array",
  "Next": "map" 
},
"map": {
  "Type": "Map",
  "ItemsPath": "$.array",
  "Parameters": {
    "payload.$": 
      "$$.Map.Item.Value"
  },
  "Iterator": {
    "StartAt": "process",
    "States": {
      "process": {
        "Type": "Task",
        "Resource": "arn:proc",
        "Parameters": {
          "payload.$": 
            "$.payload"
        },
        "End": true
      } 
    } 
  },
  "ResultPath": "$.res",
  "End": true 
}
\end{minted}
      \caption{AWS Step Functions}
      \label{lst:loop-aws}
    \end{subfigure}
    \caption{Workflow invoking function \emph{process} in parallel, with inputs from zero to three and results written to \emph{res}.}
    \label{lst:loop-aws-gcp}
\end{figure}

The most important change is the programming model, affecting the implementation of the workflows, with unknown implications to workflow performance, an important property for developers~\cite{wen2021}.
As the different implementations are all provider-specific, moving workflows from one platform to another is complicated, causing vendor lock-in~\cite{sampe2021}. 
Azure uses the programming model of Durable Functions~\cite{durable-functions-semantics}, where the workflow definition is encoded within a regular program structure of an orchestrator.
The graph of functions is expressed using a mainstream programming language such as Python, as seen in the example of mapping the elements of input \emph{values} array to invocations of the \emph{process} function (Figure~\ref{lst:loop-azure}).
The computation model is built on top of stateless \emph{activity} and stateful \emph{entity} functions.
On the other hand, developers need to define their workflow using a state machine on Google Cloud Workflows and AWS Step Functions.
The workflow consists of states representing computations and transitions connecting them.
The main states include function invocations, while supplementary states encode control flow.
State languages defined with a syntax based on JSON and YAML files can be limited, verbose, and consequently difficult to debug, with missing tool support for testing and debugging already being a problem for developers~\cite{wen2022software,LEITNER2019340}. 
The example implementations in Figure~\ref{lst:loop-aws-gcp} demonstrate how simple code snippets can become much more verbose when compared to a native implementation of orchestrator. 
In Durable Functions, implementing the same behavior requires less work and
the single-source implementation is more readable and easier to debug.
However, the static form of a state machine gives the cloud provider deep knowledge of the functions executed and their order, allowing for optimizations.

The programming model also has an impact on the billing system. 
In addition to the cost of executing functions within a workflow, cloud providers charge users for workflow orchestration.
In Azure, users have to pay for the duration of the orchestration function.
In AWS and Google Cloud, users are charged per each transition of the state machine.
Table~\ref{tab:platforms:pricing} shows an overview. Note that we have to estimate the orchestration cost on Azure as billing is at the granularity of complete workflows only. 

With the different platform-specific implications of implementing a workflow, it is difficult for developers to predict workflow costs on a given platform. 
To efficiently support them during the development of serverless workflows,
we need a higher-level construct for workflows to abstract away the differences between platforms, enabling evaluation of the same workflow on different platforms and therefore facilitating informed decisions about the right platform. 

\begin{table}
\small
    \centering
    \begin{tabular}{lrrr}
    \specialrule{.1em}{0em}{0em} 
        Platform & Compute time & Invocation & Orchestration \\\hline
        AWS &  \$0.0000167/GBs & \$0.20 per 1M & \$0.025 \\
        GCP &  \$0.0000025/GBs & \$0.40 per 1M & \makecell{\$0.01 (internal), \\ \$0.025 (external)} \\
        Azure & \$0.000016/GBs & \$0.20 per 1M & \$0.000355 \\
    \specialrule{.1em}{0em}{0em} 
    \end{tabular}
    \caption{Pricing according to vendors' documentation~\cite{aws-workflow-pricing, azure-pricing, gcp-workflow-pricing, cloud-functions-pricing, aws-lambda-pricing}. Orchestration per 1000 transitions.}
    \label{tab:platforms:pricing}
\end{table}

\subsection{Workflow Nets} 
\label{sec:background:workflow-nets}

\begin{figure}
    \centering
    \includegraphics[width=.9\linewidth]{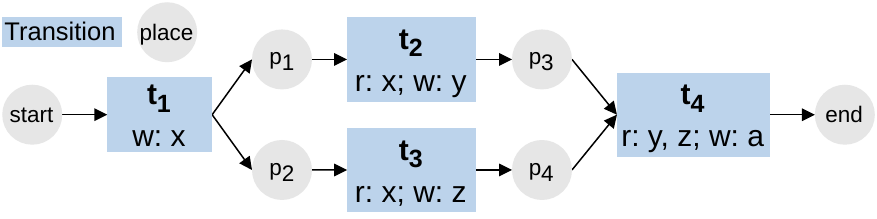}
    \caption{WFD-net with transitions $T = \{ t_1, t_2, t_3, t_4 \}$ and places $P = \{ p_1, p_2, p_3, p_4, start, end \}$}
    \label{fig:background:wfd}
\end{figure}

We base our model on workflow nets with data (WFD-nets)~\cite{wfd-nets-2}. They are an extension of Petri nets, usually used for business workflows. 
Basing the model on Petri Nets is only one possibility among alternatives such as state machines. We opt for Petri Nets due to their advantages as modeling formalism, such as their graphical nature, formal semantics, and analysis defined.
Petri nets~\cite{petrinets} describe the flow of information and control in concurrent and asynchronous systems. 
A Petri net is a triple $T = \langle P, T, F \rangle$ consisting of places P, a finite set of transitions T, and a set of arcs $F \subseteq (P \times T ) \cup (T \times P)$. 
It is a workflow net \textit{iff} there is a single source place start without incoming arcs, a single sink place without outgoing arcs, and every node is on a path from source to sink~\cite{workflow-nets-with-data}. WFD-nets~\cite{workflow-nets-with-data} are a tuple $\langle P, T, F, D, r, w, d, grd\rangle$, consisting of a Petri Net $N =\langle P, T, F\rangle$ and additionally containing a set $D$ of data elements on top as well as read, write, and destroy operations on these data elements. Moreover, the guarding function $grd: T \rightarrow G_D$ can assign guards to transitions. We show an example in Figure~\ref{fig:background:wfd} where $t_1$ writes data to $x$, while $t_2$ and $t_3$ read from x. 
Dynamic system properties are modeled using tokens that are routed through the net. 
A transition is enabled if tokens are in all its input places $\bullet t = \{p | (p, t) \in F \}$. When it fires, it removes the token(s) from its input place(s) and routes them to its output place(s) $t\bullet = \{p | (t, p) \in F \}$. In our example, $t_1$ will be enabled if there is a token in $start$ and put tokens to $p_1$ and $p_2$, which will enable $t_2$ and $t_3$. 

The platforms orchestrating serverless workflows that impose time limits on execution and schedule functions. Moreover, it is important to model how in- and output data is passed between functions. Modeling both of these is currently not supported by WFD-nets. 

\section{Serverless Workflows Model} \label{sec:model}

We define a model for serverless workflows that allows developers to implement and analyze a workflow application independent of the platform it will run on, alleviating provider lock-in. 
The model should encode the control flow and task parallelism, and 
clearly display the flow of data between functions, aiding developers in detecting scalability bottlenecks and errors, e.g., inconsistent or missing data. 
Therefore, we define our model on top of WFD-nets~\cite{workflow-nets-with-data} (cf. Section~\ref{sec:background:workflow-nets}) and extend them to be able to express the orchestration by the platform and how data is passed between functions.

\begin{figure}[t]
    \centering
    \includegraphics[width=\linewidth, keepaspectratio]{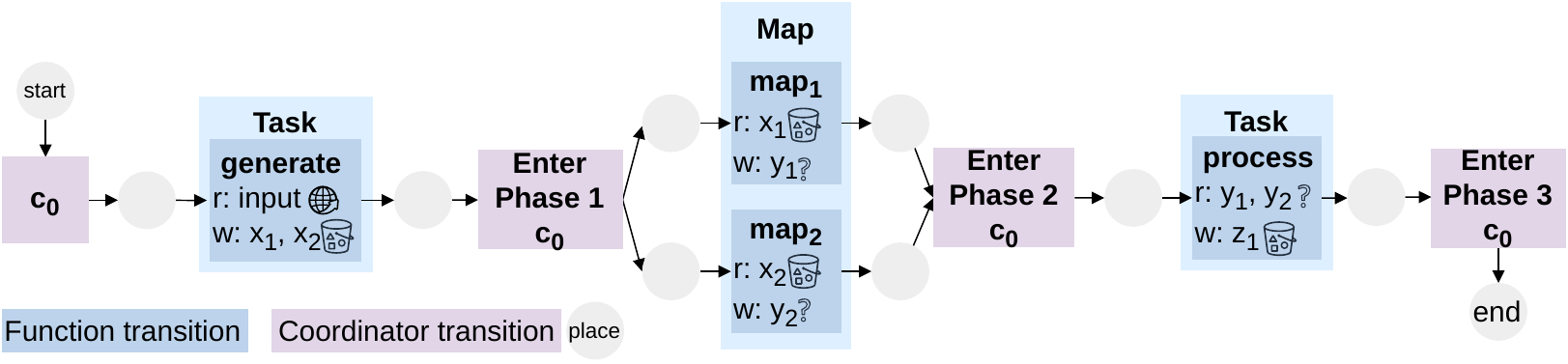}
    \caption{Workflow using our model based on WFD-nets.}
    \label{fig:model:func-group-coordinator-transitions}
\end{figure}

\subsection{Transitions} 
\label{sec:model:transitions}

The set of transitions $T$ is composed of two types, the coordinators $C$ and serverless functions $SF$, $T = C \cup SF $. Figure~\ref{fig:model:func-group-coordinator-transitions} shows an example with $C = \langle c_0, Enter Phase 1 c_0, Enter Phase 2 c_1\rangle$ and $SF = \langle generate, map_1, map_2, process\rangle$.

A \textit{function} transition $sf \in SF$ represents the execution of a serverless function. 
All function transitions that can run in parallel without any precedence dependencies and their immediate predecessor and successor places make up a workflow phase. 
There are different possible token routing constructs within one phase of the workflow: 
A \emph{task} phase is a sequential routing, consisting of one function transition only. 
For parallel routing, there are two alternatives:
First, a \emph{parallel} phase can consist of any number of sub-phases that will be executed concurrently. 
Second, the \emph{map} phase: Similar to the parallel phase, it can consist of any number of sub-phases, but each sub-phase is executed concurrently on different elements of an input array. 
Figure~\ref{fig:model:func-group-coordinator-transitions} shows an example: The \texttt{map} functions compute $y_i = \texttt{map}\left( x_i\right)$ simultaneously for all $i$.
A \emph{switch} phase uses conditional routing based on values of data by annotating guarding functions to transitions.

The first transition of a workflow in our model is always a \textit{Coordinator} $c \in C$ that initializes the workflow and schedules functions for execution. 
Additional coordinator transitions take place between phases, meaning that the coordinator awaits the termination of the currently running functions and afterwards schedules the functions of the next phase, explicitly modeling the orchestration of the workflow by the platform. 
For readability, we do not show the coordinator transitions when they can be skipped while preserving the control flow between function transitions, i.e., whenever a sequential phase is the next phase. This is because the sequential function already serves the purpose of the AND-join otherwise realized by the coordinator transition. 
In Figure~\ref{fig:model:func-group-coordinator-transitions}, this means we can leave out all coordinator transitions after the initial $c_0$ transition.

\subsection{Resource Annotations}

Data labeling functions indicate the required inputs and provided outputs of a transition. 
However, for the performance of serverless workflows, it is important to know where the data resides and how it is provided. Therefore, we extend the notation of WFD-nets by annotating how the data is passed using the following resource annotations: 

\begin{itemize}[leftmargin=*]
    \item \caseicon{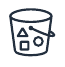} \textbf{Object storage.} Data is saved in cloud storage in the same region. While providing high capacity, it suffers from limited I/O bandwidth and high latency.
    \item \caseicon{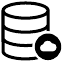} \textbf{NoSQL.} \eurosys{Data stored in NoSQL key-value storage provides low-latency data storage.}
    \item \caseicon{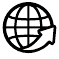} \textbf{Invocation Payload.} Protocols such as HTTP and gRPC can transfer small input data. However, the exact size limit is subject to the protocol and platform. 
    \item \caseicon{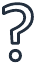} \textbf{Transparent.} The type of transmission used when returning a payload is up to the provider and can change given the payload size.
    \item \caseicon{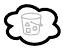} \textbf{Reference.} Some functions only need the reference to an object in the object storage rather than the object itself. 
\end{itemize}

\eurosys{Formally, we define the set of resource annotations $A = \{ o, n, p, t, r \} $ as additional element of the tuple of a WFD-net, with $o$ representing data passing via the object storage, $n$ via NoSQL, $p$ via the invocation payload, $t$ transparently, and $r$ via reference. 
We define the corresponding resource annotation functions for reading and writing data as $ra$ and $rw$ as follows and also add them to the tuple of a WFD-net:}
\begin{equation*}
\text{ra}: \{ (t, d) \in T \times D \mid d \in r(t) \} \to A     
\end{equation*}
\begin{equation*}
\text{rw}: \{ (t, d) \in T \times D \mid d \in w(t) \} \to A
\end{equation*}

\eurosys{This means that each pair of a transition and a data element $(t, d)$, with $d$ being read or written by $t$, respectively, is assigned a resource annotation $a \in A$.}
\eurosys{By adding resource annotations, we do not change the behavior of the WFD-net. However, we enable checking the consistency of data accesses, for example, if the same data object is written and read using the same resource annotation.}

\noindent
We annotate data location in workflows using the respective icon and show an example in Figure \ref{fig:model:func-group-coordinator-transitions}. The function \texttt{generate} receives a payload via an invocation payload and stores its output on the object storage. The \texttt{map} functions each receive an element of the array, process it, and return their resulting elements $y_1$ and $y_2$ through a protocol decided by the cloud provider. Once both \texttt{map} functions have returned, the \texttt{process} function receives $y_1$ and $y_2$ as input and, finally, uploads the final result $z$ of the workflow to the object storage.
\section{Workflows Benchmark Suite}
\label{sec:impl}

We now present the design and implementation of \toolname{}\footnote{
An extended definition and discussion of benchmarks can be found in the Master thesis~\cite{laurinthesis}.
}.
To enable reliable and fair comparison of various workflow platforms, we need to execute the same benchmark implementation on many platforms. 
However, the platforms exhibit vast differences in the programming model and API of their workflow services (Section~\ref{sec:platforms}).
Thus, we define a platform-agnostic workflow definition (Section~\ref{sec:impl:definition}) based on our workflow model (Section~\ref{sec:model}). 
Then, we propose platform-specific generators that transcribe workflows to the respective proprietary definition of the desired platform (Section~\ref{sec:impl:generators}).
We add the workflow representation and implementation to a serverless benchmark suite (Section~\ref{sec:impl:benchmark-suite}).

\subsection{Platform-Agnostic Workflow Definition}
\label{sec:impl:definition}

Our workflow model encodes the application as Petri Net (cf. Section~\ref{sec:model}). \eurosys{To define workflows in \toolname{} conforming to our model, we use a JSON syntax.}
Every phase has a \texttt{type}, relating to one of the available routing constructs (cf. Section~\ref{sec:model:transitions}). 
Coordinator transitions encode the order of phases, represented by the \texttt{next} field of phases that describes the consecutive step in the workflow. \eurosys{The \texttt{next} field refers to the phase name to be executed after, and the workflow terminates if this field is not set.} 
Each phase receives the output payload of the previous function as input. \eurosys{This means that function implementations need to conform to the resource annotations as defined in the workflow model and download and upload data as needed accordingly.}
We encode the different phases as follows:

\begin{figure}[t]

    \begin{subfigure}{0.49\linewidth}

      \begin{subfigure}[b]{\linewidth}

        \footnotesize
\begin{minted}{json}
"compute_phase": {
  "type": "task",
  "func_name": "compute"
}
\end{minted}
        \caption{\eurosys{\emph{Task} Statement.}}
        \label{lst:statement-task}

      \end{subfigure}
      \begin{subfigure}[b]{\linewidth}
        \footnotesize
\begin{minted}{json}
"process_names": {
  "type": "map",
  "array": "customers",
  "root": "shorten",
  "next": "list_emails",
  "states": { 
    "shorten": {
      "type": "task",
      "func_name": "short"
    } 
  } 
}
\end{minted}
        \caption{\emph{Map} Statement.}
        \label{lst:statement-map}
      \end{subfigure}

    \end{subfigure}
    \hfill 
    \begin{subfigure}{0.49\linewidth}

      \footnotesize
\begin{minted}{json}
"root": "generate_phase",
"states": {
  "generate_phase": {
    "type": "task",
    "func_name": "generate",
    "next": "map_phase" },
  "map_phase": {
    "type": "map",
    "array": "x",
    "root": "map",
    "next": "process_phase",
    "states": {
      "map": { 
        "type": "task",
        "func_name": "map" } } },
  "process_phase": {
    "type": "task",
    "func_name": "process" 
  } 
}
\end{minted}
      \caption{\eurosys{Workflow from Figure~\ref{fig:model:func-group-coordinator-transitions}.}}
      \label{lst:complete-workflow}

    \end{subfigure}
    \caption{Workflow definition language: a portable specification of control-flow and data dependencies.}
    \label{lst:statements}
\end{figure}

\noindent
\textit{Task.} A \texttt{task} executes a single serverless function\eurosys{, constituting a sequential routing}. Listing~\ref{lst:statement-task} shows an example with the \emph{compute\_phase} executing the function \emph{compute}.

\noindent
\textit{Map.} 
The \texttt{map} phase \eurosys{is a parallel routing construct and} concurrently executes the given states one after another on each element of the given array and returns an array again. 
The phase can define \texttt{common\_parameters} from the running variable that will be passed in addition to the array element. 
Listing \ref{lst:statement-map} shows an example with the \texttt{process\_names} phase: for each element of \texttt{customers}, the function \texttt{short} is executed concurrently. 
Only after all functions have terminated, the coordinator will transition to the next phase, which in this case is \texttt{list\_emails}.

\noindent
\textit{Loop.} The \texttt{loop} phase is similar to \texttt{map} but traverses the given input array sequentially. Thus, loop encodes tasks that cannot be parallelized due to existing dependencies. 

\noindent
\textit{Repeat.} A \texttt{repeat} phase executes a function a given number of times. 
This syntactic sugar eases modeling a chain of tasks. 

\noindent
\textit{Switch.}
The \texttt{switch} phase \eurosys{is a conditional routing, deciding the next phase dynamically at runtime based on the given condition. The different \texttt{cases} are evaluated after another, with the first one fulfilling the condition being executed.}  

\noindent
\textit{Parallel.}
This higher-level phase \eurosys{corresponds to a parallel routing and} executes sub-work\-flows, consisting of any of the phases, concurrently. 

\eurosys{We show an example of a complete workflow definition in Listing~\ref{lst:complete-workflow}, encoding the same workflow as shown in Figure~\ref{fig:model:func-group-coordinator-transitions}.
The \texttt{root} entry specifies the name of the phase that should be executed first, in this case the \texttt{generate\_phase}. The \texttt{states} entry then contains all phases of the workflow. 
As mentioned above, each phase receives the output payload of the previous function as input. 
Therefore, the data movement is encapsulated in the functions and not controlled by the workflow orchestration. 
Only in the case of the \texttt{map} phase, we explicitly specify which array is used for distributing its elements to single functions. The level of parallelism is then decided dynamically at runtime depending on the size of the given array.
}

\subsection{Platform-Specific Transcription} \label{sec:impl:generators}

We map the six phases building a serverless workflow to different features of the modeling language on each platform. \eurosys{We evaluate the overhead introduced by necessary adaptations to platforms in Section~\ref{sec:evaluation-model}.}

\subsubsection{AWS}
The most notable difficulty when transcribing our definition to the state machine definition of AWS Step Functions is the \texttt{loop} phase. 
Step Functions do not inherently support sequential array iteration. 
Their official documentation suggests using an additional serverless function that iterates over a given range~\cite{step-functions-iterator}, which is inefficient. 
Thus, we use the AWS \texttt{map} state and configure it to traverse the given array sequentially, yielding the semantics of a \texttt{loop}.
A downside of this approach is that the input to each function is the same, i.e., consecutively executed functions can observe the results of computations of their predecessors only if uploaded to the object storage.

\subsubsection{Google Cloud}
Google Cloud Workflows do not natively support a \texttt{task} type.
Instead, the recommended approach for invoking Cloud Functions~\cite{google-cloud-functions} is to create a state performing a POST request and providing the trigger URL of the desired function as input.
However, this requires additional states for each \texttt{task} and \texttt{map} to parse the HTTP response of a function and assign results. 
Moreover, the parallel \texttt{map} execution accepts only other workflows and not states, which requires creating another sub-workflow, even if it contains only a single function to be invoked.
Finally, there is no mechanism for passing additional arguments to a \texttt{map} function, which is necessary for us to track measurements.
As a workaround, the input array is zipped together with an array consisting of the additional parameter passed by the benchmarking infrastructure.

\subsubsection{Azure}
Azure uses the dynamic model of Durable Functions instead of state machines.
There, we upload our workflow definition together with the function code. 
The user-provided orchestrator parses the definition as input, decodes our definition, and executes it by spawning new function executions.

\subsection{Benchmark Suite}
\label{sec:impl:benchmark-suite}

We follow standard design practices to build a new benchmark suite: it should be relevant, extensible, easy to use, and reproducible
~\cite{v.Kistowski:2015:BB:2668930.2688819,Binnig:2009:WTT:1594156.1594168,benchmarking,copik2021sebs}.
Our suite is relevant as we include applications representing a variety of workloads in the industry and academia (Section~\ref{sec:benchmarks}).
The implementation is based on an abstract workflow definition and can be extended to new platforms 
by implementing a single interface that transcribes our model definition to the new platform. 
To fulfill the two remaining criteria, we build our implementation upon SeBS~\cite{copik2021sebs}, an established benchmark suite for FaaS:
Benchmarks must be easy to deploy and execute to ensure their self-validation~\cite{v.Kistowski:2015:BB:2668930.2688819}.
Integration into a maintained and up-to-date platform helps integrating new developments of serverless platforms continuously and avoids pushing this task to the end user. 
\toolname{} is multi-platform, supports automatic deployment of functions to the cloud, and integrates with services like storage and cloud logging, allowing developers to focus on the actual implementation rather than specifics of cloud providers, which can be time-consuming~\cite{chatley2020,sashko2024}.

Serverless functions need 
cloud storage to access data and retain state across invocations.
To that end, SeBS automatically manages object storage instances and provides functions with a multi-cloud API.
To create realistic workflow representations of web applications,
we need to support low-latency data stores other than object storage.
We chose NoSQL key-value storage for this task and extended SeBS with a high-level interface for creating, modifying, retrieving, and deleting items. The interface supports a partition and an optional sorting key.
Each benchmark function can use multiple tables managed by the benchmark suite.
We map the tables to DynamoDB on AWS, CosmosDB on Microsoft Azure, 
and Firestore in Datastore mode on Google Cloud.

We collect timestamps for \textit{start} and \textit{end} of each function, its \textit{requestID}, and a \textit{containerID} to detect container reuse by using the temporary filesystem and global variables. 
The runtime of a phase is defined by the \textit{start} of its earliest function and the \textit{end} of the latest one. 
All collected values are sent to a Redis~\cite{redis} instance deployed in the same cloud region.
We chose an in-memory cache as it provides sub-millisecond latencies, reducing the risk of distorting the performance measurements. 

\begin{figure}[t]
    \centering
    \includegraphics[width=\linewidth, keepaspectratio]{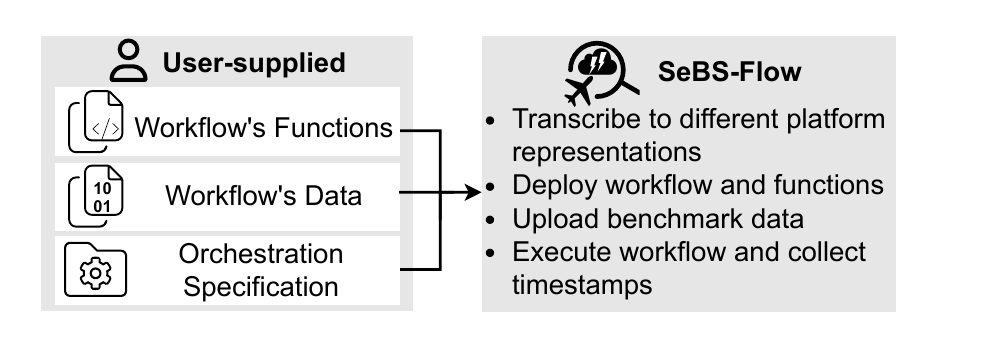}
    \caption{\cameraready{Process of executing a workflow using \toolname{}.}}
    \label{fig:impl:process}
\end{figure}

\cameraready{
To implement a workflow using our model, a user has to provide the following (cf. Figure~\ref{fig:impl:process}): 
First, the implementation of the workflow’s functions in a language of their choice. Our workflow model is independent of the actual benchmark implementation. We can work with any language supported in the cloud, with currently supported Python, Node.js, C++, and Java through SeBS. 
Second, any data used as input to the workflow. 
Third, the specification how the functions should be orchestrated using our platform-agnostic workflow definition in JSON (cf. \autoref{sec:impl:definition}). 
\toolname{} takes both as input and deploys the workflow with the functions to the respective cloud the user chooses, transcribing the workflow to their platform-specific representations by traversing the JSON file. This is transparent to the user and fully automated.}

\section{Benchmark Applications}
\label{sec:benchmarks}

In \toolname{}, we implement six benchmarks covering real-life workloads.
Also, we implement four microbenchmarks used in the evaluation: function chain, object storage performance, parallel invocations (Section~\ref{sec:eval:performance:overhead}), and selfish detour (Section~\ref{sec:eval:performance:discrepancy}).
The selected benchmarks cover various domains that use workflows (Table~\ref{tab:benchmarks-features}),
and correspond to previous findings on the characterization of workflow use cases~\cite{eismann2020review,eismann2021review} regarding control-flow, number of functions, parallel invocations of the same functions, longer runtimes\cameraready{, and workload sizes}: 
\cameraready{We include the sequential TripBooking benchmark (50\% of workflows), four benchmarks using less than ten different functions (72\% of workflows), five benchmarks involving parallel invocations of the same function (52\% of workflows), and the 1000Genome workflow contains functions with a runtime of over a minute (25\% of workflows). Moreover, the analysis by \citet{eismann2021review} classifies 72\% of serverless workflows as “small,” consisting of 2 to 10 functions, 23\% as “medium” with 11 to 1000 functions, and only 4\% as “large” with more than 1000 functions. With our workloads, four of six applications can be classified as “small” and two more as “medium.”}
\cameraready{
Furthermore, the analysis of traces from production workloads on Azure Durable Functions~\cite{wisefuse} also showed that 40\% of workflows are sequential, most workflows do not use more than ten functions, the median number of different functions in a workflow is three, and the median workflow execution time is 5.6 seconds, confirming the representativeness of our applications.}
We visualize only one of the benchmarks here, but provide figures for the other benchmarks in the supplementary material.

\begin{table}
    \centering
    \begin{adjustbox}{width=\linewidth}
    \begin{tabular}{lcccccr}
    \specialrule{.1em}{0em}{0em} 
        Benchmark & \#functions & Parallelism & Critical path & Download [MB] & Upload [MB] \\\hline
        Video & 4 & 2 & 3 & 238.83 & 7.48 \\
        Trip Booking & 7 & 1 & 4/7 & 0.0 & 0.0 \\
        MapReduce & 9 & 5 & 4 & 0.02 & 0.04 \\
        ExCamera & 16 & 5 & 6 & 302.07 & 17.49 \\
        ML & 3 & 2 & 2 & 7.82 & 3.91 \\
        1000Genome & 19 & 12 & 4 & 273.54 & 3.47 \\
    \specialrule{.1em}{0em}{0em} 
    \end{tabular}
    \end{adjustbox}
    \caption{Key features of different benchmarks.}
    \label{tab:benchmarks-features}
\end{table}

\paragraph{Video Analysis} \label{sec:benchmark:video}

\begin{figure}
    \includegraphics[width=\linewidth]{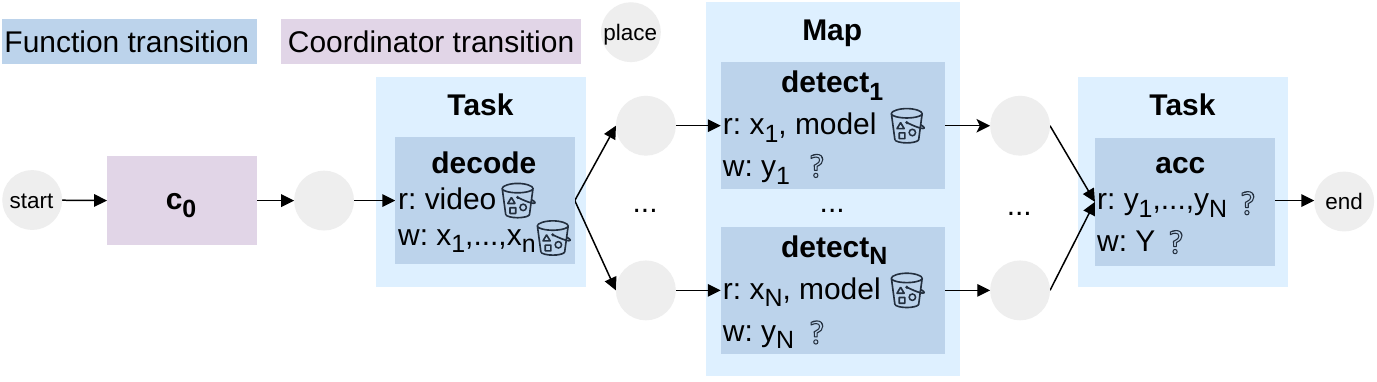}
    \caption{The Video Analysis benchmark.}
    \label{fig:benchmark:vid}
\end{figure}

The benchmark detects objects in a video, and parallelizes the sequential benchmark in vSwarm~\cite{vswarm} (Figure~\ref{fig:benchmark:vid}).
Functions decode video frames and apply the Faster R-CNN model~\cite{faster-rcnn}.
The \texttt{decode} function first downloads the video, decodes $F$ frames, and then uploads $N=\lceil \frac{F}{B}\rceil$ batches of size $B$. 
$N$ parallel \texttt{detect} functions compute $Y_i$, 
all detections with confidence $p > 0.5$. Finally, detections are accumulated in \texttt{acc}, returning the final payload $Y$. 
We used $F=10$ frames and batch size $B=5$, yielding two parallel functions in the map phase.

\paragraph{Trip Booking}
The benchmark represents web applications, and it mocks a common example of reserving a hotel, car rental, and flight~\cite{sagapattern1,sagapattern2}.
The workflow is a pipeline of functions mocking the reservation system by storing trip data in a shared NoSQL database.
It implements the SAGA pattern of long-running transactions~\cite{10.1145/38714.38742} where a failure triggers the reversal of prior changes.
For testing, we simulate failure in the last \emph{confirm} function, which is followed by three consecutive functions to reverse the booking.

\paragraph{MapReduce}
We base our example on prior implementations~\cite{vswarm, towardsdatascience} and perform the standard problem of word counting.
First, the \texttt{split} function partitions the input text into $N$ batches. $N$ parallel \texttt{map} functions count how often each word occurs in their text chunk next. 
Next, \texttt{shuffle} flattens the resulting array $Y_i | i < M$. Finally, $M$ reducers count the total occurrences of their respective word in parallel, yielding $Z_i$. 
The benchmark has two parameters: the number of mapping functions $N$, and the total number of words $W$.
We set $N=3$ and $W=5000$, containing $M = 5$ different words.
MapReduce frameworks typically execute fully in parallel. However, the available workflow primitives necessitate the \texttt{shuffle} function, not relying on the array $Y_i$ itself but flattening it to enable the desired level of parallelism in reduce.

\paragraph{ExCamera}
ExCamera~\cite{fouladi2017-original-excamera} uses interdependent video-processing tasks to encode videos in parallel. A video with $M$ total frames is processed in chunks of $N$ frames by $\frac{M}{N} = T$ parallel functions. 
First, each frame is encoded, yielding one key frame and $N-1$ interframes.
Decode decodes all $N$ frames again, calculating the final state. 
The final state from the first frame of the chunk is used for reencoding the other frames, resulting in one final state and $N-2$ interframes.
We derive our implementation from the original description of ExCamera~\cite{fouladi2017-original-excamera} and the available implementation~\cite{fouladi2019-gg-excamera}. 
We use $M = 30$ total frames and a chunk size of $N = 6$, resulting in five parallel functions.

\paragraph{Machine Learning}

This workload represents a typical training pipeline: 
It starts with \texttt{gen} generating a dataset, with the number of samples $N$ and the number of features $M$ as input.
Then, we \texttt{train} $K$ different classifiers $C_{i}$ in parallel.
We generate $N=500$ samples and $M=1024$ features, and train $K=2$ classifiers: a Support Vector Machine~\cite{svc}, and a Random Forest~\cite{randomforest}, creating two concurrent functions. 

\paragraph{1000Genomes}
This is a scientific workflow that identifies mutational overlaps using data from the 1000 Genomes project~\cite{1000genome-project}. 
It consists of five tasks and three phases: 
First, $N$ \texttt{individuals} functions parse the data for their chunk of the input file of size $M$ and then upload their results to the cloud storage. 
While \texttt{individuals\_merge} merges the results to one, \texttt{sifting} computes the Sorting Intolerant from Tolerant (SIFT) scores. 
In the last phase, \texttt{mu\-ta\-tion\_over\-lap} measures the overlap in Single Nucleotide Polymorphisms (SNP) variants and \texttt{frequency} measures the frequency of mutation overlapping, both by population $P$. 
The benchmark has the number of lines as input 
$M$, number of parallel \texttt{individuals} functions $N$, and number of populations $P$ as input variables.
We use $M = 1250$ lines, $N = 5$ parallel individuals function, and $P = 6$ populations. 

\section{Evaluation of Workflow Model}
\label{sec:evaluation-model}

By reviewing existing literature on serverless workflows, we evaluate whether our model is general enough to express applications of workflows 
and if our transcription 
to the platform-specific representations 
adds overhead compared to the native implementation. 
We do so by using the meta-search engine Google Scholar to find peer-reviewed publications containing the keywords \textit{cloud, orchestration}, and \textit{serverless workflow} or \textit{serverless DAG}. 
We exclude papers that are not in English, do not use a workflow benchmark, or are published before 2017, the year of the first serverless workflows in the cloud.
This results in 72 papers analyzed papers~(cf. Table~\ref{tab:table_intro}, p.~\pageref{tab:table_intro} for their categorization). 
We provide the complete list of papers and analysis results in the supplementary material.

\subsection{Expressiveness of our Model}

We analyze the workflow benchmarks used in the literature and evaluate whether our model can represent the control flow within the workflows without adding unnecessary dependencies between their tasks. 
Out of the 72 papers, 14 did not provide sufficient detail on the workflows used and their dependencies to judge if we can express them. 
In two papers, benchmarks are not presentable by our model, as they introduce new programming models to support communication between functions and load-balanced orchestration.
Benchmarks used in three more papers can be modeled but not transcribed to platform-specific representations~(Section~\ref{sec:impl}).
For two of them, cloud platforms are the limitations, such as ending the workflow as a result of a switch state (not possible on AWS) and using multi-stage inputs, i.e., using the output of a previously executed function as input without passing it to the functions invoked in-between.
\eurosys{The third one uses a \textit{switch} state requiring two conditions to be true.} While we do not support transcribing \eurosys{this currently, transcription} the \textit{switch} state requiring two conditions to be true, it can be easily added to the implementation.
We fully support modeling and transcribing the workflows described in 53 of the 58 analyzed papers.
Therefore, we conclude that our model does not have general limitations \eurosys{within the scope of programming models not allowing for communication between functions and using orchestration based on dynamic characteristics of the system,} and developers can use it to model and execute their workflows.

\subsection{Overhead of our Model}

To check if our model and transcription \eurosys{(cf. Sec.~\ref{sec:impl:generators})} create overhead compared to a native implementation, we evaluate available benchmark implementations used in the analyzed papers and compare them to our transcription of their workflows.
Only 10 of the 72 papers include an artifact containing workflow implementations or show their implementation as part of the paper for any of the platforms we support. 
None of them uses Google Cloud Workflows. 
In total, we find eleven AWS Step Functions state machines.
One of them uses the \textit{AND} choice type. \eurosys{We currently do not transcribe this choice type and are therefore not able to generate the same state machine. However, if we would add the transcription, the resulting state machine would look similar.} 
Another one adds \emph{fail} and \emph{success} states before ending the workflow, which only introduces overhead as compared to just ending the workflow.
The other nine state machines use the same states with the same parameters in the same order as the state machines we transcribe, except for the fact that they specify each parameter explicitly as part of the state machine while we wrap them within a single \textit{payload} entry\eurosys{, which does not affect the overhead}. 
Four of the papers provide implementations for a total of six workflows using Azure Durable Functions. 
While one paper only provides an implementation using entity functions, the other five workflow implementations use activities to orchestrate tasks similar to our transcription. 
Since we must parse the platform-independent representation within the orchestrator, we could introduce an overhead.
However, the evaluation of the 1000Genome benchmark, the benchmark with the most functions, shows that the average duration of the orchestrator function is only 13.6 milliseconds\cameraready{, with the workflow's median runtime being 3757.55 seconds}. 
We conclude that \toolname{} does not introduce noteworthy overhead in the workflows compared to their native implementation, enabling developers to obtain realistic performance results for their workflows.

\subsection{Threats to Validity}

We used only one query to find relevant works, bearing the risk of missing results. 
We mitigated this by evaluating different queries beforehand, evaluating the relevance of papers found, and checking if the results included relevant papers we knew \eurosys{as a gold standard~\cite{dieste2009developing}}. 
Regarding external validity, we found only a limited number of artifacts to evaluate the overhead, 
with none available that uses GC Workflows. 
While our transcription follows best practices and tutorials as provided by the cloud providers and matches the artifacts we found, usage in other projects could differ. 
\section{Evaluation of Cloud Services}
\label{sec:evaluation}

We use \toolname{} to evaluate three major cloud workflow services -- AWS Step Functions, Google Cloud Workflows, and Azure \cameraready{Durable Functions} -- providing developers valuable insights regarding their suitability for different workloads. We investigate the following research questions:

\begin{enumerate}[label=\textbf{RQ\arabic*},leftmargin=*]
    \item\label{RQ1} What are the runtime differences between platforms?
    \item\label{RQ2} What causes runtime and stability differences?
        \begin{enumerate}[leftmargin=10pt,label=\textbf{RQ2.\arabic*}]
            \item\label{RQ2.1} \eurosys{What causes overheads in the orchestration?} 
            \item\label{RQ2.2} What causes variations in the critical path?
        \end{enumerate}
    \item\label{RQ3} How well can serverless workflow orchestration support scientific workflows? 
    \item\label{RQ4} How does the pricing compare between platforms? 
    \item\label{RQ5} How did the performance and stability of the platforms evolve over time?
 \end{enumerate}

\subsection{Methodology}
We deploy benchmarks \cameraready{and resources used by them} on Azure to the \emph{europe-west} region, on AWS to \emph{us-east-1},
and on Google Cloud to \emph{us-east1}.
We use the lowest common memory configuration that successfully executes the workflow on AWS and Google Cloud, at least 256 MB for computational functions and 128 MB for simple web applications.
We invoke the application benchmarks in \textit{burst} mode, triggering 30 executions at once and accepting all successful workflow executions, as other work suggests that most serverless applications have potentially bursty workloads~\cite{eismann2021review}. 
We check how often we should repeat experiments by computing non-parametric confidence intervals on the measurements of the MapReduce benchmark and aim at being in a 5\% interval of the median using a 95\% confidence interval. 
For the burst mode with 30 executions triggered at once, this results in 1, 1, and 6 repetitions on AWS, GCP, and Azure, respectively. 
We opt to execute all experiments 180 times. 
However, we could only obtain 30 executions of the 1000Genome benchmark on Azure due to frequent timeout issues. 
Benchmarks use the serverless object storage and NoSQL database on each platform.

\subsection{RQ1: Runtime Differences among Platforms} \label{sec:eval:performance}

\begin{figure}
\begin{subfigure}{0.49\linewidth}
\centering
\includegraphics[width=\linewidth]{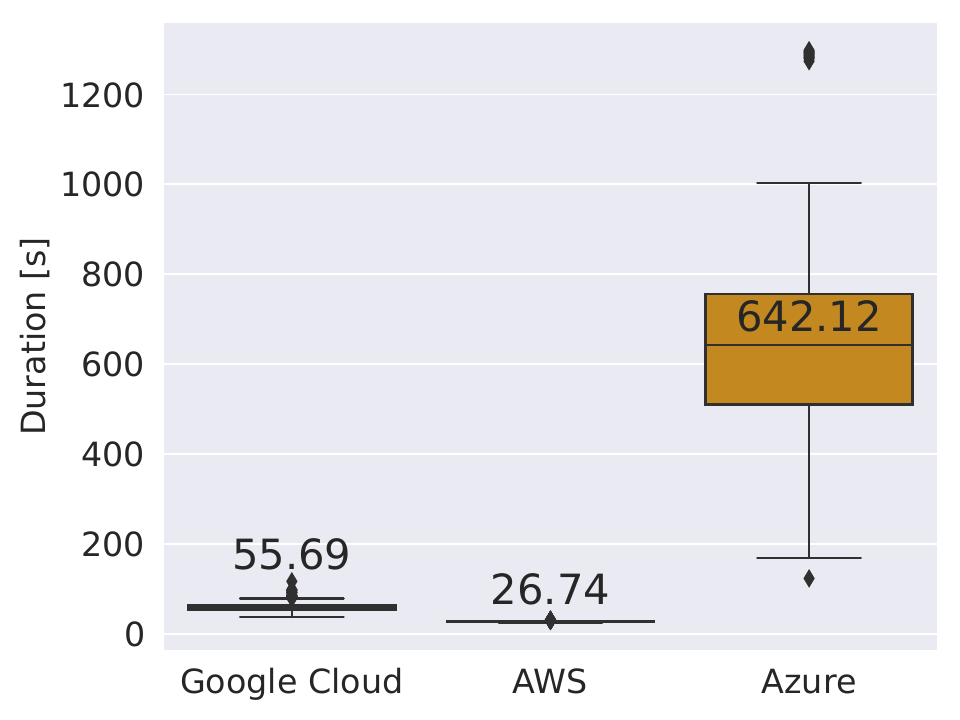}
\caption{Video Analysis\ifARXIV, 2048MB.\fi}
\label{fig:eval:performance:video-analysis}
\end{subfigure}\hfill
\begin{subfigure}{0.49\linewidth}
\centering
\includegraphics[width=\linewidth]{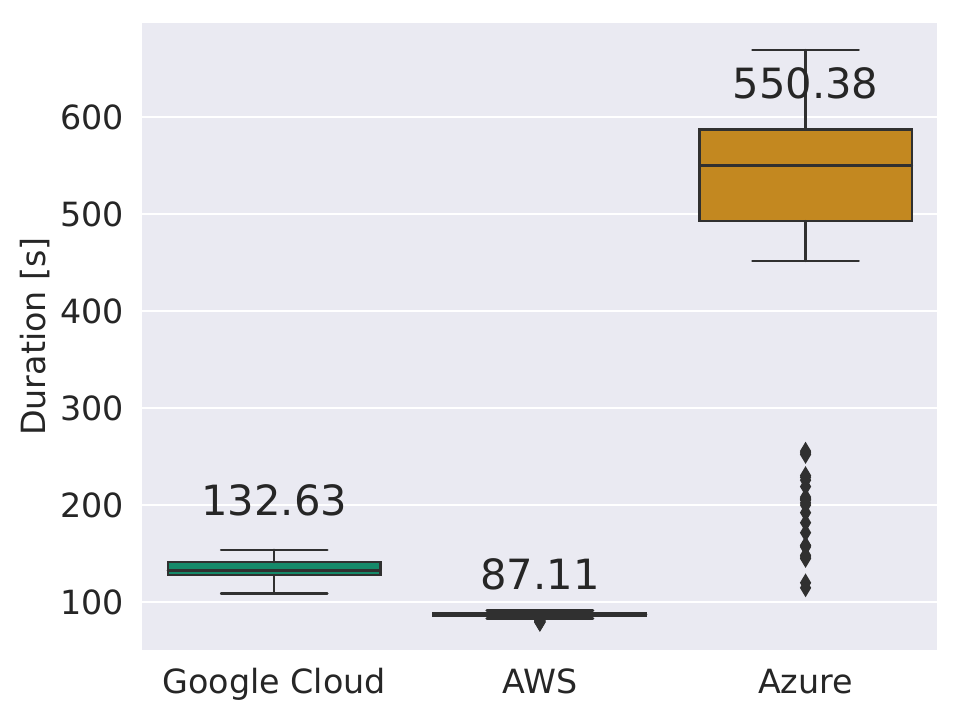}
\caption{ExCamera\ifARXIV, 256MB.\fi}
\label{fig:eval:performance:excamera}
\end{subfigure}\hfill
\begin{subfigure}{0.49\linewidth}
\centering
\includegraphics[width=\linewidth]{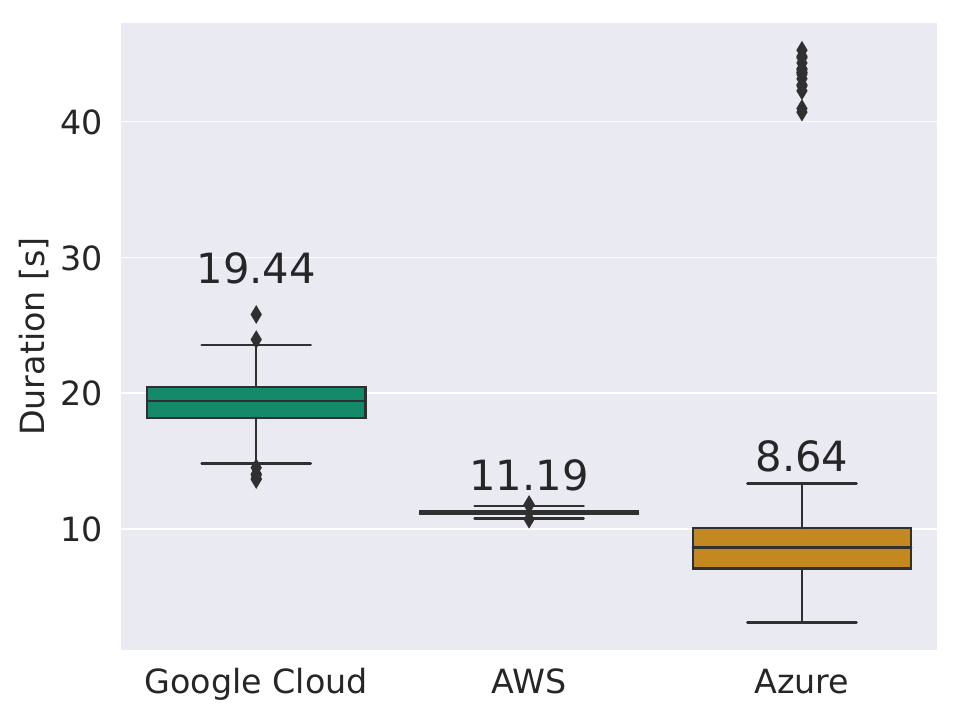}
\caption{MapReduce\ifARXIV, 256MB.\fi}
\label{fig:eval:performance:mapreduce}
\end{subfigure}\hfill
\begin{subfigure}{0.49\linewidth}
\centering
\includegraphics[width=\linewidth]{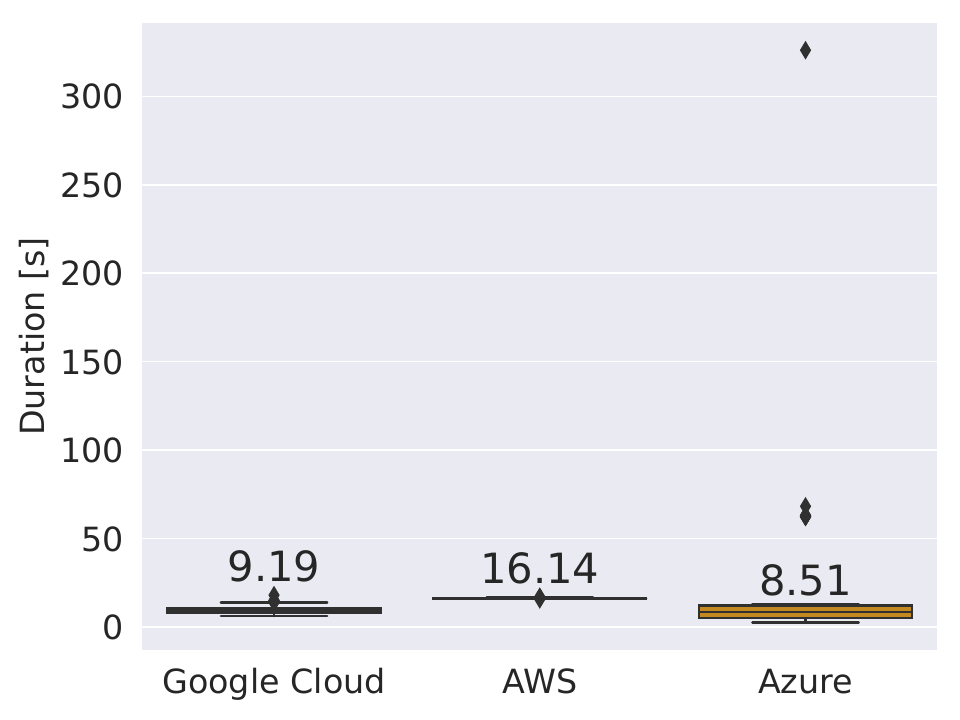}
\caption{Trip Booking\ifARXIV, 128MB.\fi}
\label{fig:eval:performance:trip-booking}
\end{subfigure}\hfill
\begin{subfigure}{0.49\linewidth}
\centering
\includegraphics[width=\linewidth]{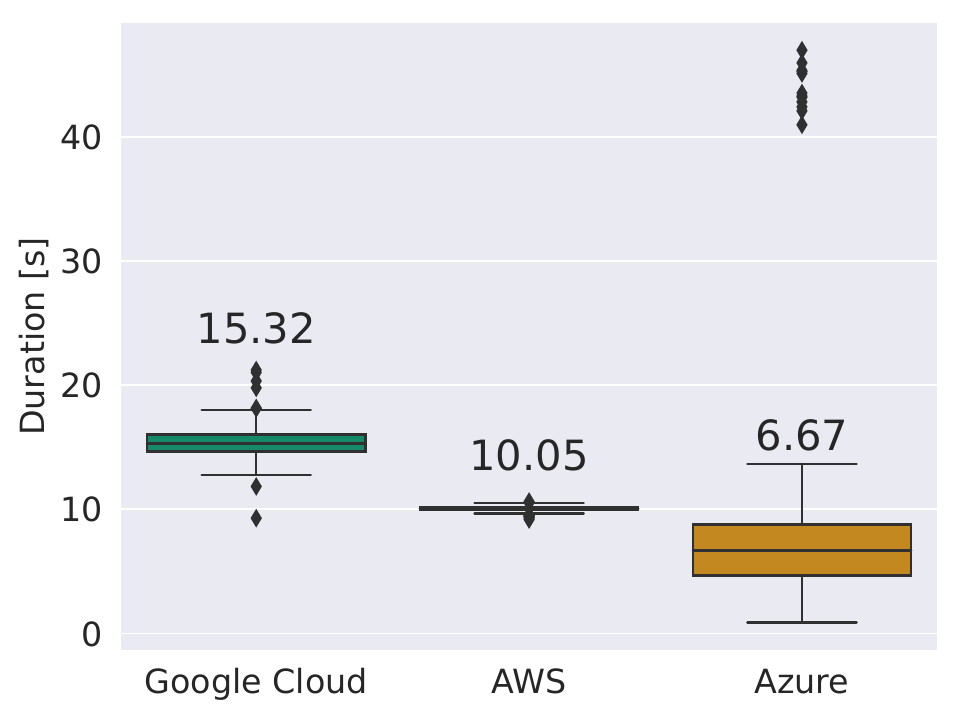}
\caption{\small{Machine Learning\ifARXIV, 1024MB.\fi}}
\label{fig:eval:performance:ml}
\end{subfigure}\hfill
\begin{subfigure}{0.49\linewidth}
\centering
\includegraphics[width=\linewidth]{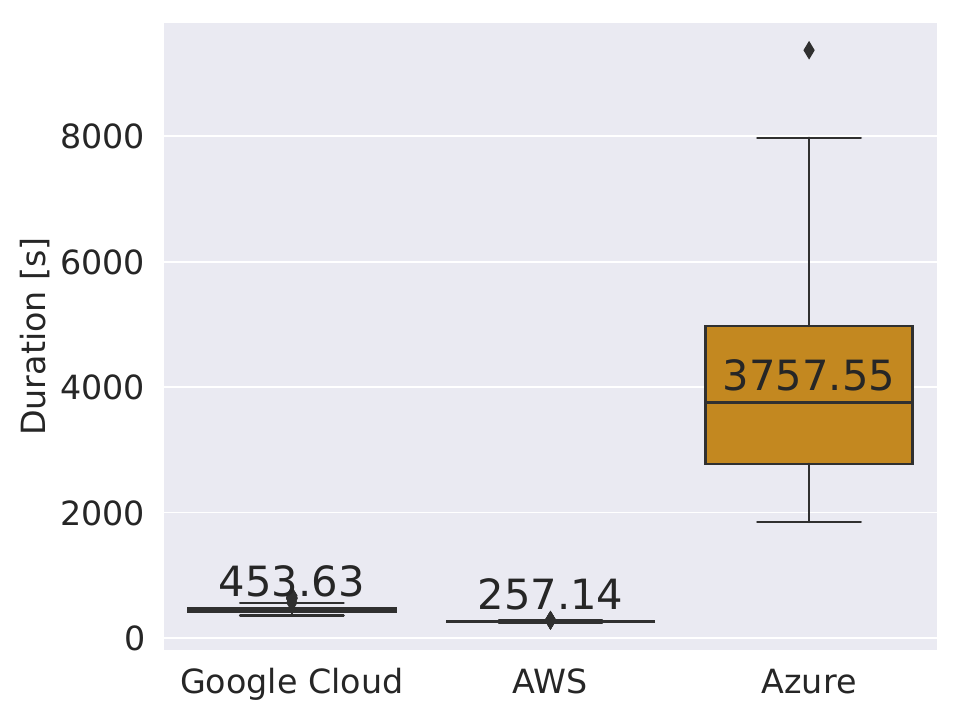}
\caption{1000Genome\ifARXIV, 2048MB.\fi}
\label{fig:eval:performance:1000genome}
\end{subfigure}\hfill
\caption{Runtime of benchmark applications on AWS Step Functions, GC Workflows, and Azure Durable\ifARXIV, \textit{burst} invocations\fi.}
\label{fig:eval:performance}
\end{figure}

We compare the runtime of each benchmark on the selected platforms.
We calculate the runtime by subtracting the first \textit{start} timestamp from the last \textit{end} timestamp. 
The results in Figure~\ref{fig:eval:performance} do not yield a single fastest platform among all our benchmarks.
AWS is the fastest platform for three out of six benchmarks while performing relatively well for the other three. 
While Google Cloud's performance is comparable to AWS, it is 1.55-1.97x slower on three benchmarks. 
While Azure Durable functions perform very well, e.g., on MapReduce and Machine Learning,
they are the slowest platform for Video Analysis, ExCamera, and the 1000Genome benchmark.
For Trip Booking, Azure achieves the best median performance but suffers from large outliers.
We investigate the potential causes of slowdown in the next section.
All platforms demonstrate variable performance, with Azure showing the largest variance.

\subsection{RQ2: Causes for Runtime and Stability Differences}
\begin{figure}
\begin{subfigure}{0.49\linewidth}
\centering
\includegraphics[width=\linewidth]{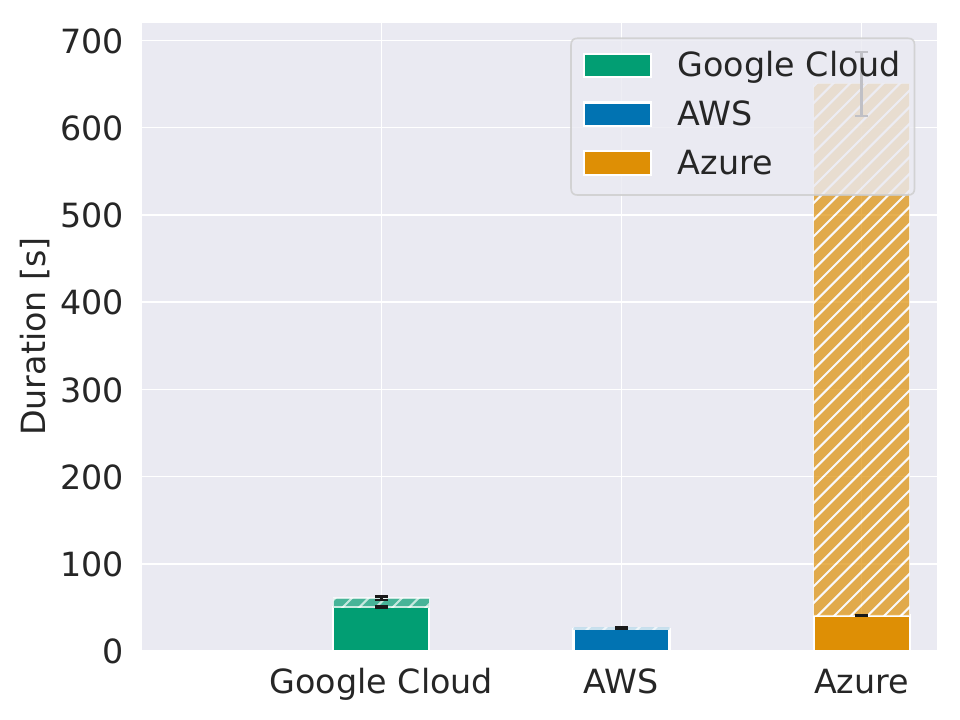}
\caption{Video Analysis, 2048MB.}
\label{ffig:eval:overhead:video-analysis}
\end{subfigure}\hfill
\begin{subfigure}{0.49\linewidth}
\centering
\includegraphics[width=\linewidth]{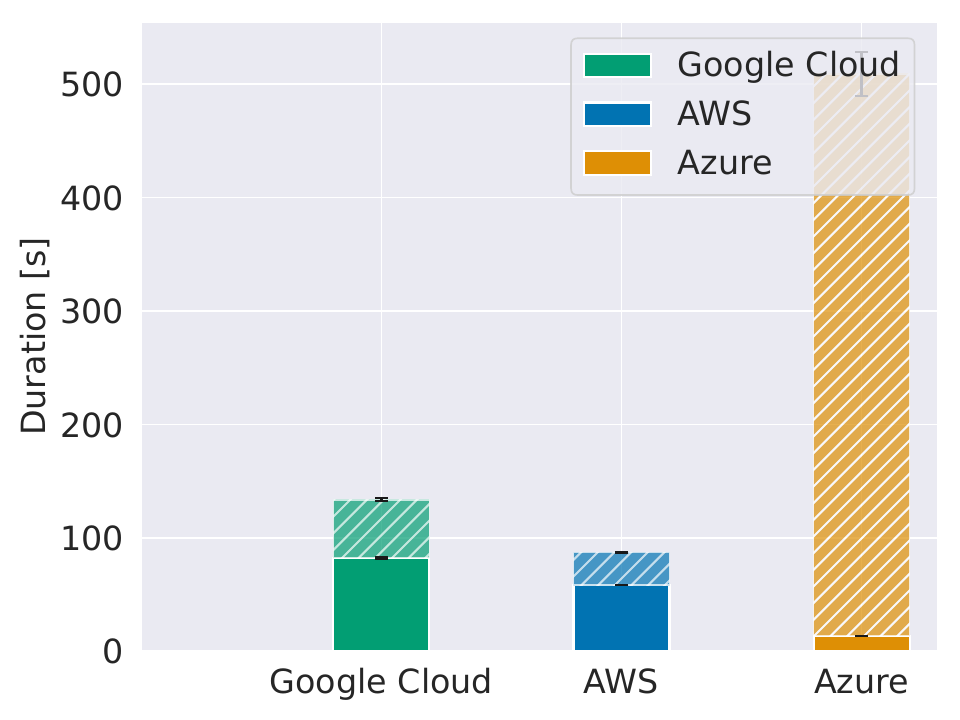}
\caption{ExCamera, 256MB.}
\label{fig:eval:overhead:mapreduce}
\end{subfigure}\hfill
\begin{subfigure}{0.49\linewidth}
\centering
\includegraphics[width=\linewidth]{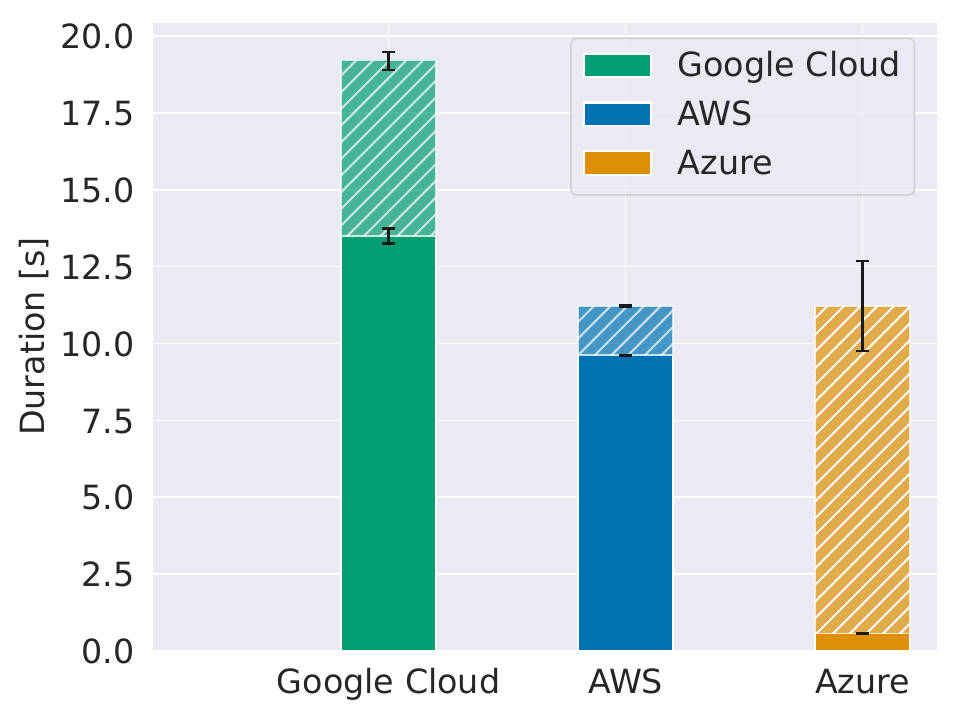}
\caption{MapReduce, 256MB.}
\label{fig:eval:overhead:excamera}
\end{subfigure}\hfill
\begin{subfigure}{0.49\linewidth}
\centering
\includegraphics[width=\linewidth]{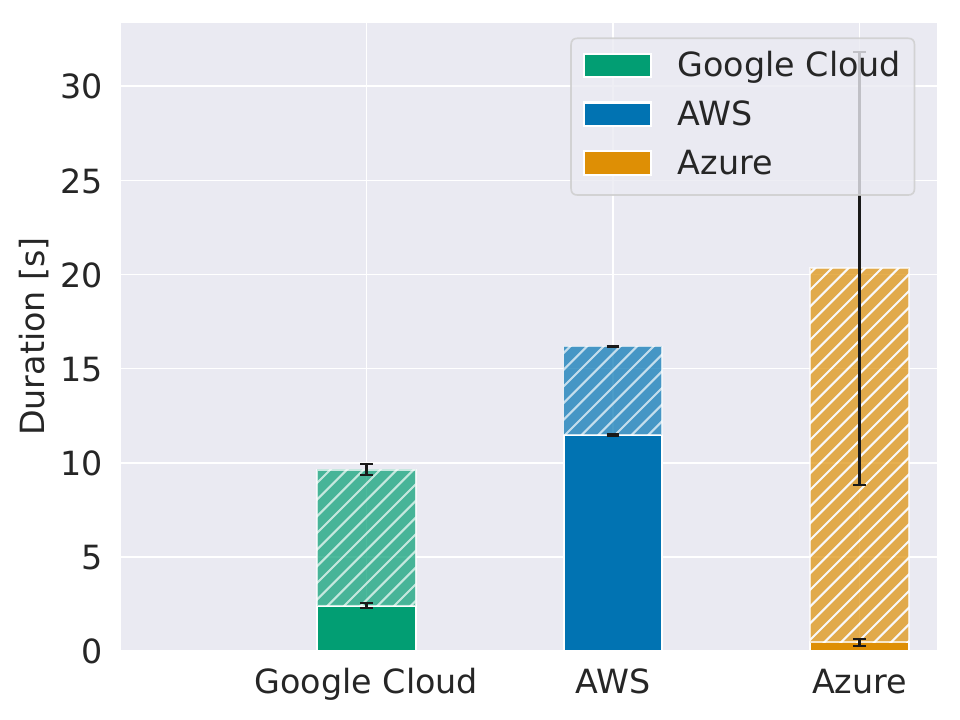}
\caption{Trip Booking, 128MB.}
\end{subfigure}\hfill
\begin{subfigure}{0.49\linewidth}
\centering
\includegraphics[width=\linewidth]{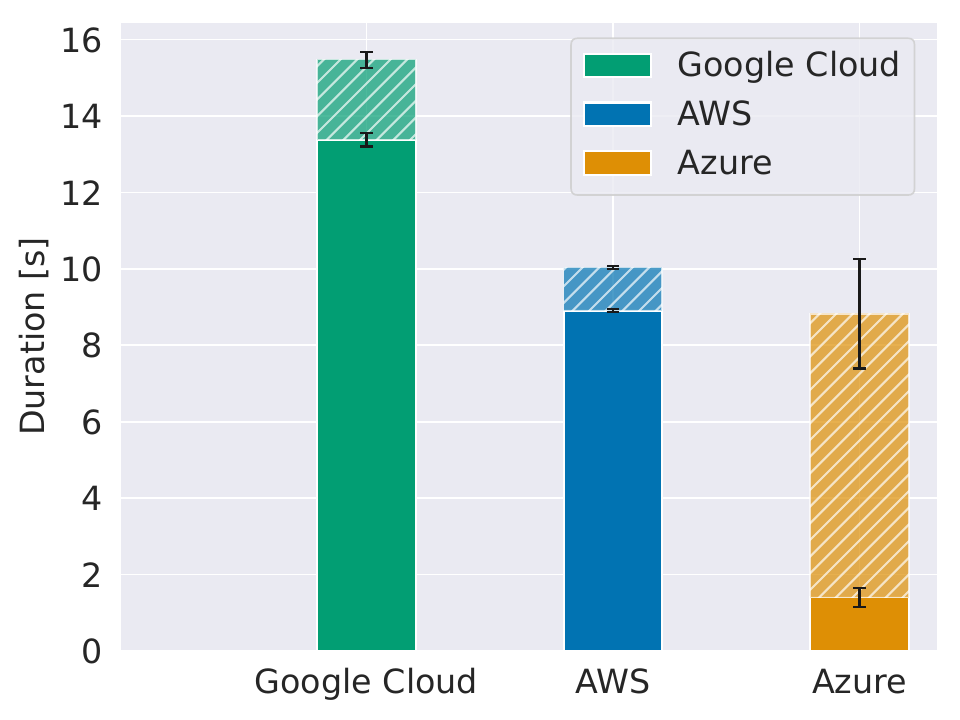}
\caption{Machine Learning, 1024MB.}
\label{fig:eval:overhead:ml}
\end{subfigure}\hfill
\begin{subfigure}{0.49\linewidth}
\centering
\includegraphics[width=\linewidth]{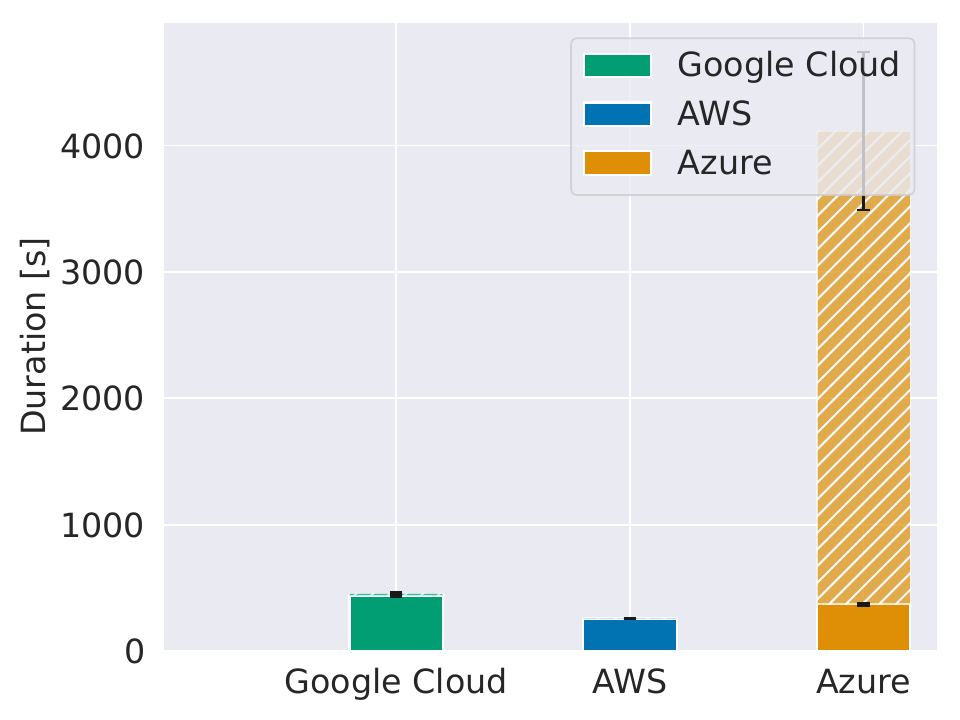}
\caption{1000Genome, 2048MB.}
\label{fig:eval:overhead:1000genome}
\end{subfigure}
\caption{Critical path (opaque) and overhead (hatched) of different benchmarks on considered platforms\ifARXIV, \textit{burst} invocations\fi.}
\label{fig:eval:overhead}
\end{figure}

According to our results, AWS and Google Cloud provide a performance-reliable workflow service, whereas the variability is considerably higher on Azure.
Thus, we split the runtime into two components \eurosys{to investigate the reasons behind this}: the critical path $T_C$, computed as the sum of all states' maximum runtime within one phase, and the overhead $T_O$ caused by the scheduling and data movement conducted by the cloud workflow service.
We calculate the overhead $T_O$ by subtracting the critical path $T_C$ from the total runtime. 
Figure~\ref{fig:eval:overhead} shows the critical path and overhead for all benchmarks.
Azure's runtime is dominated by \eurosys{highly variable} scheduling overhead: For example, the overhead of the ExCamera benchmark is, on average, 495.5s, more than 36$\times$ as long as its critical path of 13.5s. The ML benchmark incurs the least overhead of 5$\times$ the length of its critical path. 
Also, Azure's critical path is very fast across all benchmarks, demonstrating the fastest critical path for ExCamera, MapReduce, and Machine Learning. 
Google Cloud, however, has the slowest critical path throughout the entire benchmark suite. 
In summary, orchestration overhead causes long runtimes and performance variances on Azure. 
For AWS and Google Cloud, however, the critical path varies. 
\eurosys{We therefore explore different causes for the differing behavior of the cloud platforms in the following Sections~\ref{sec:eval:performance:overhead} and~\ref{sec:eval:performance:discrepancy}}.

\subsubsection{RQ2.1 Sources of Overhead} \label{sec:eval:performance:overhead}

We analyze three common sources of overhead: object storage I/O, parallel schedule, and function return payload. 

\begin{figure}
\begin{subfigure}{0.49\linewidth}
    \centering
    \includegraphics[width=\linewidth, keepaspectratio]{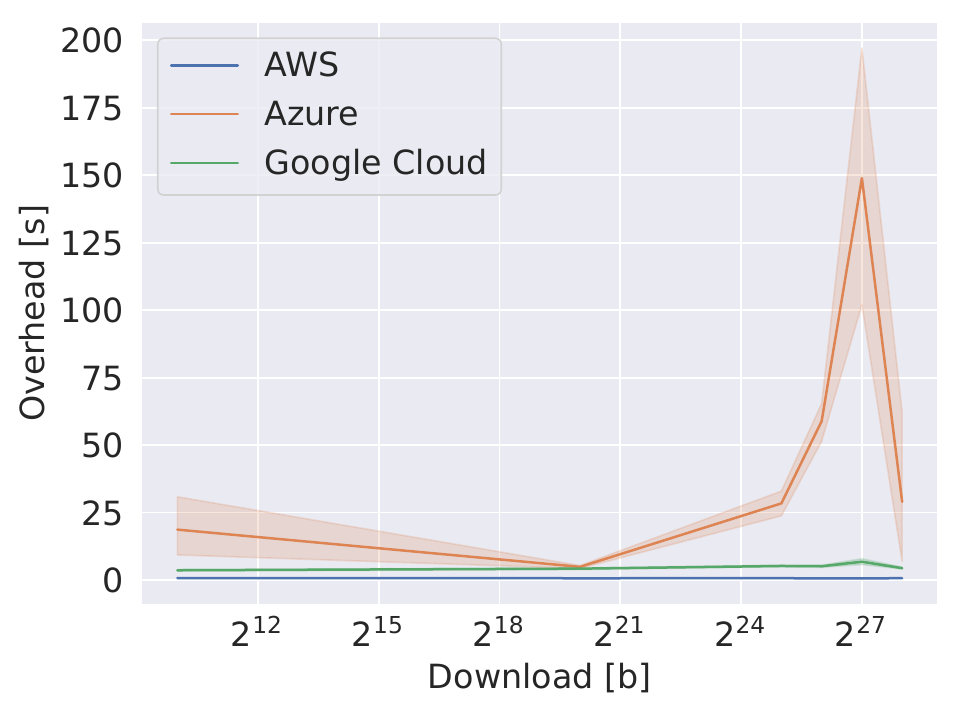}
    \caption{Overhead of storage I/O, 20 functions, $2^{10} \leq D \leq 2^{28}$, 512MB, \textit{burst} invocations.}
    \label{fig:eval:overhead:parallel-download}
\end{subfigure}
\hfill
\begin{subfigure}{0.49\linewidth}
    \centering    
    \includegraphics[width=\linewidth, keepaspectratio]{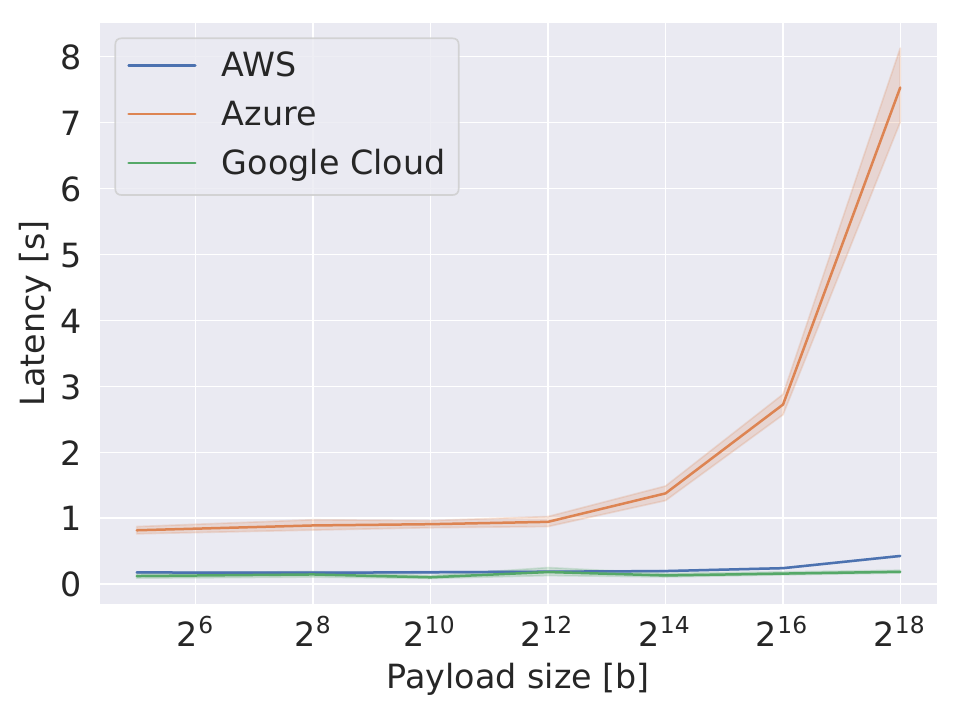}
    \caption{Invocation latency, \eurosys{chain of 10 functions}, \ifARXIV\fi $2^5 \leq M < 2^{18}$, 256MB, \textit{warm} invocations.}
    \label{fig:eval:overhead:function-chain}
\end{subfigure}
\caption{Analysis of different sources of overhead.}
\end{figure}

\paragraph{Cloud Storage I/O}
The data downloaded from the object storage differs between benchmarks (Table~\ref{tab:benchmarks-features}, p.~\pageref{tab:benchmarks-features}), with hundreds of megabytes in ExCamera, 1000Genomes, and Video Analysis.
These benchmarks experience the highest relative overhead of 36.7$\times$, 10$\times$, and 14.95$\times$ their critical paths on Azure. 
To verify that this correlation is indeed causation, we execute a microbenchmark evaluating the cloud storage I/O performance.
We invoke 20 functions in parallel where each attempts to download a file of size $D$ from the storage. 
Figure~\ref{fig:eval:overhead:parallel-download} shows that the overhead remains stagnant for AWS at around one second and nearly stagnant on Google Cloud, increasing a bit for downloads larger than 1MB.
On Azure, however, we observe an overhead of almost 149 and 4.9 seconds for 128 and 1 MB files, respectively.
%
\eurosys{Therefore, data downloads} can account for a significant part of the large overhead measured on Azure Durable.

\begin{figure}
\begin{subfigure}{0.322\linewidth}
    \centering
    \includegraphics[width=\linewidth, keepaspectratio]{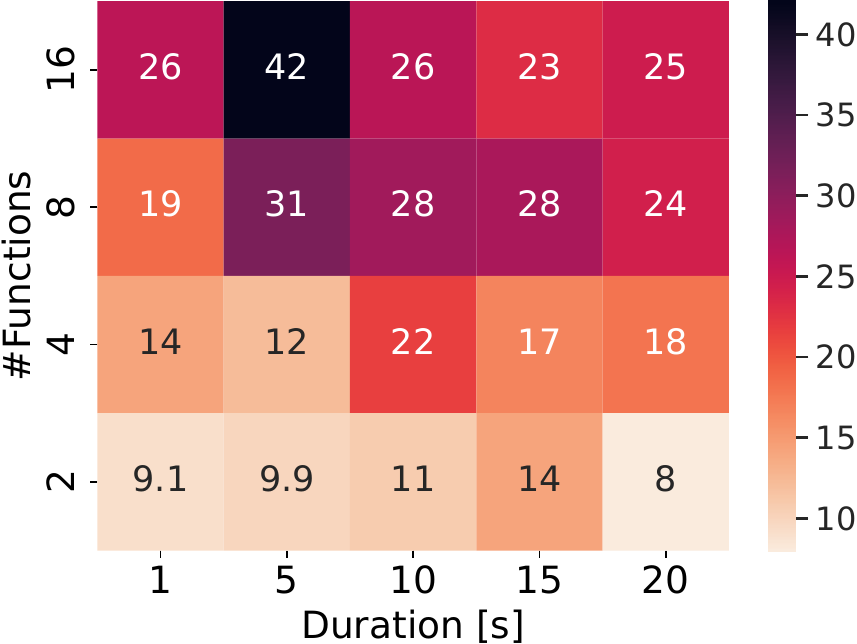}
    \caption{Azure.}
    \label{fig:eval:overhead:parallelsleep:azure}
\end{subfigure}\hfill
\begin{subfigure}{0.322\linewidth}
    \centering
    \includegraphics[width=\linewidth, keepaspectratio]{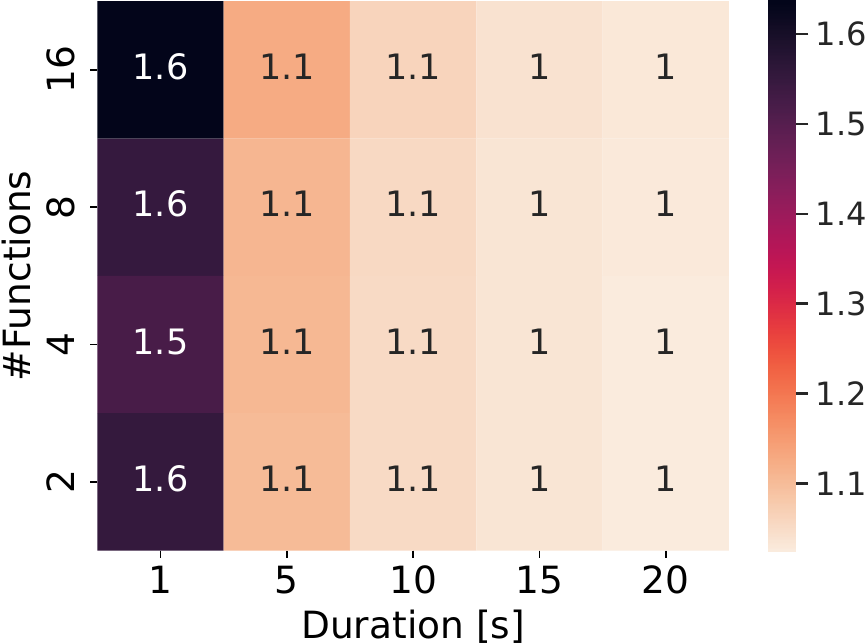}
    \caption{AWS.}
    \label{fig:eval:overhead:parallelsleep:aws}
\end{subfigure}\hfill
\begin{subfigure}{0.322\linewidth}
    \centering
    \includegraphics[width=\linewidth, keepaspectratio]{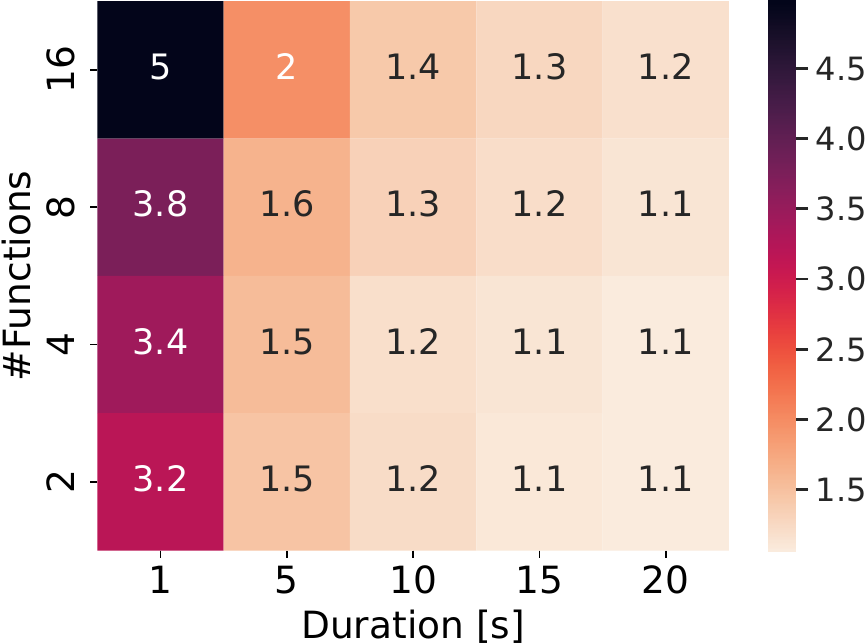}
    \caption{ Google Cloud.}
    \label{fig:eval:overhead:parallelsleep:gcp}
\end{subfigure}
\caption{The overhead of parallel sleep microbenchmark, $2 \leq N \leq 16$, $1 \leq T \leq 20$, 256MB, \textit{burst} invocations.}
\label{fig:eval:overhead:parallelsleep}
\end{figure}

\paragraph{Parallel Scheduling}
Another potential source of overhead are parallel invocations within a benchmark: Benchmarks with the highest degree of parallelism -- ExCamera and 1000Genomes -- show the largest overheads of Azure.
We test this by executing a microbenchmark that spawns $N$ functions in parallel, each one sleeping for $T$ seconds, and start 30 such invocations concurrently.
Figure~\ref{fig:eval:overhead:parallelsleep} shows the relative overhead of the actual runtime \eurosys{of the workflow} compared to the function execution time. 
AWS functions demonstrate modest overhead, with largest values for the shortest duration. \cameraready{The absolute overhead incurred for a certain number of parallel functions remains relatively constant, causing the relative overhead to decrease with higher function executin times.} 
GC functions present a larger \cameraready{relative} slowdown that increases with the number of parallel tasks. 
There, the system puts a cap on scaling up and reuses containers, as 30 invocations with $N = 2, T = 1$ start 60 different function containers on AWS, but only 30 on Google Cloud. 
On the other hand, Azure experiences an order of magnitude larger \cameraready{relative} overhead that increases with the parallelism factor but does not seem to be correlated to the function runtime.

\begin{figure*}
\begin{subfigure}{0.199\linewidth}
\centering
\includegraphics[width=\linewidth]{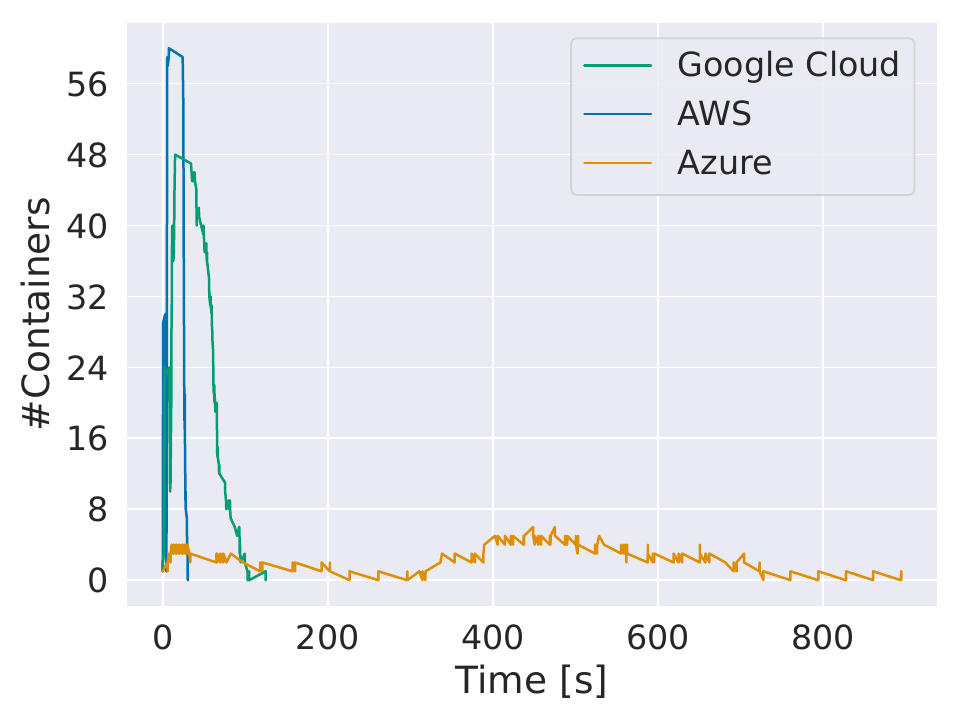}
\caption{Video Analysis}
\label{fig:eval:scalability:video-analysis}
\end{subfigure}\hfill
\begin{subfigure}{0.199\linewidth}
\centering
\includegraphics[width=\linewidth]{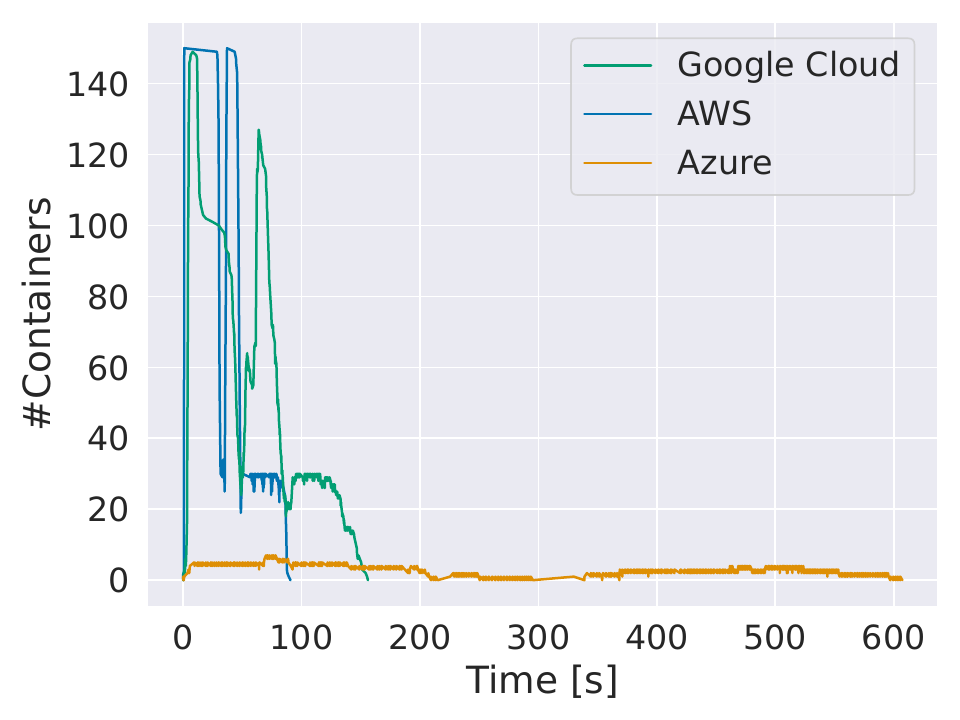}
\caption{ExCamera}
\label{fig:eval:scalability:excamera}
\end{subfigure}\hfill
\begin{subfigure}{0.199\linewidth}
\centering
\includegraphics[width=\linewidth]{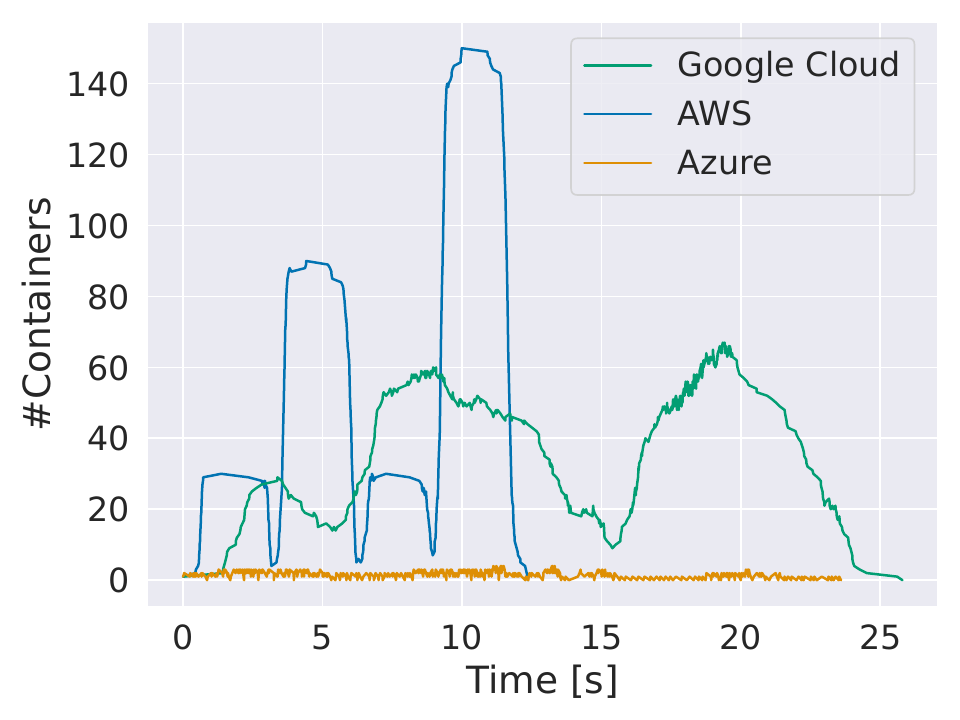}
\caption{MapReduce}
\label{fig:eval:scalability:mapreduce}
\end{subfigure}\hfill
\begin{subfigure}{0.199\linewidth}
\centering
\includegraphics[width=\linewidth]{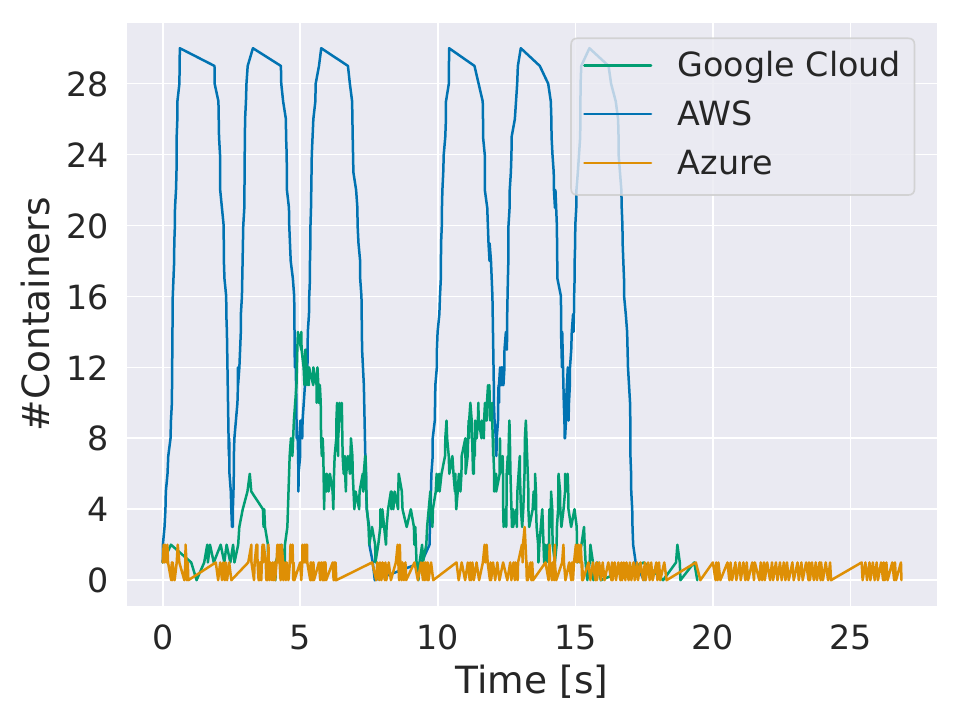}
\caption{Trip Booking}
\label{fig:eval:scalability:trip-booking}
\end{subfigure}\hfill
\begin{subfigure}{0.199\linewidth}
\centering
\includegraphics[width=\linewidth]{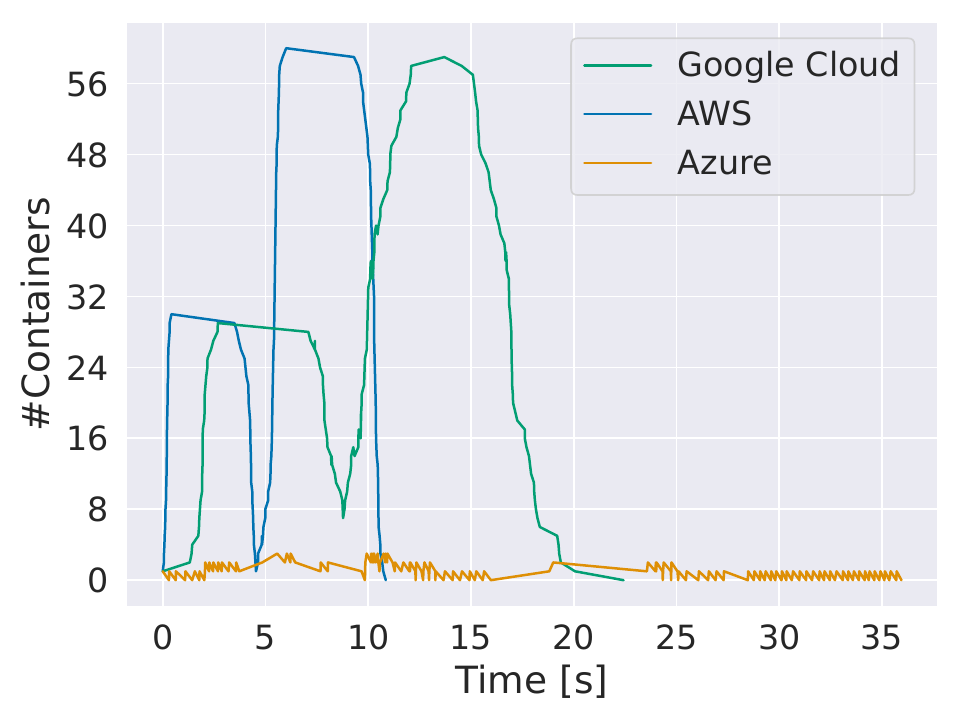}
\caption{Machine Learning}
\label{fig:eval:scalability:ml}
\end{subfigure}
\caption{Scaling profiles: the number of distinct containers used for 30 consecutive workflow invocations.}
\label{fig:eval:scalability}
\end{figure*}

To better understand the impact of limited parallel scalability on our benchmarks, we measure the number of distinct sandboxes allocated at any given time until the last function execution has terminated. 
We invoke 30 concurrent executions of workflow benchmarks and display the scaling behavior in Figure~\ref{fig:eval:scalability}.
Throughout the benchmarks, AWS and Google Cloud exhibit similar scaling behaviors, and their scale-up curves reveal the same local maxima, with phase transitions visible.
However, we can also see that AWS spins up new containers more quickly.
Azure produces a much more constant curve that remains similar throughout the benchmarks, never allocating more than 10 containers simultaneously.

\paragraph{Return Payload}
We evaluate the overhead resulting from the function return payload size.
We deploy a microbenchmark consisting of a function chain, where functions return $M$ bytes of result sent to the consecutive function, with ten functions and test varying input sizes until Google Cloud's limit. We invoke the chain 30 times simultaneously and use results from warm invocations only.
Figure~\ref{fig:eval:overhead:function-chain} shows that the latency remains constant for AWS and Google Cloud, while it increases dramatically for Azure from 16 kB, suggesting an influence of remote storage or queue.
While this may present a significant source of overhead in applications, our benchmarks do not return payloads larger than 1MB, and this overhead can only account for a part of the slowdown.

\paragraph{Conclusions}
The microbenchmarks demonstrate that a significant part of the overhead observed on Azure originates from the parallel schedules and storage I/O.
\eurosys{Moreover, the return payload can be a source of overhead on Azure for larger payloads.
To minimize overheads, workflows downloading large amounts of data, using high levels of parallelism, and high return payloads may therefore better be deployed to AWS or Google Cloud, with AWS demonstrating less overheads and better scalability across our benchmarks. }

\subsubsection{RQ2.2 Critical Path Discrepancy}
\label{sec:eval:performance:discrepancy}
The runtime of benchmarks across platforms shows that additionally to varying overhead,
the critical path of computation can be significantly different.
To understand the reasons behind this difference, we analyze how the critical path is impacted by two factors: the varying CPU allocation and frequency of cold starts.

\begin{figure*}

\begin{minipage}{0.599\linewidth}
\begin{subfigure}{0.322\linewidth}
\centering
\includegraphics[width=\linewidth, keepaspectratio]{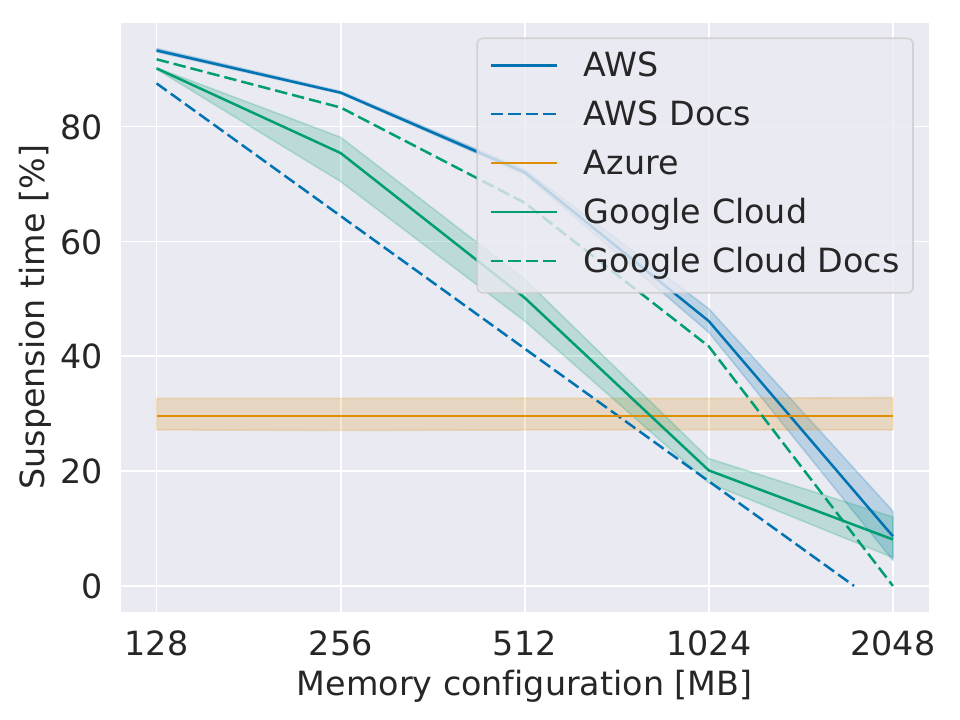}
\caption{Relative suspension, \small{$N=5000$}, \textit{warm}.}
\label{fig:eval:osnoise:selfish-detour}
\end{subfigure}
\begin{subfigure}{0.322\linewidth}
\centering
\includegraphics[width=\linewidth, keepaspectratio]{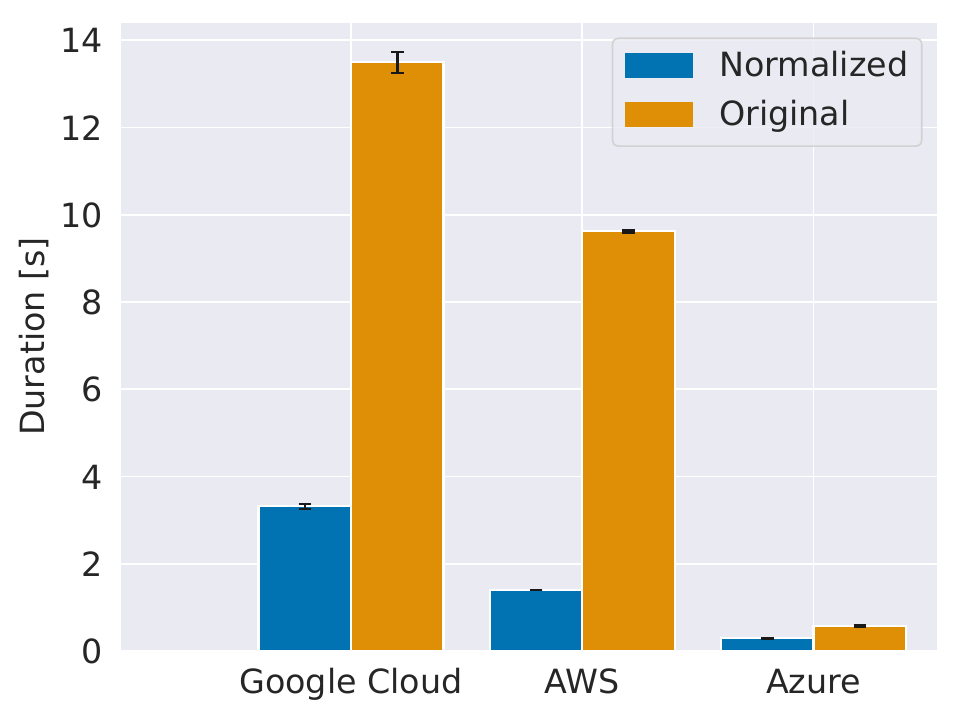}
\caption{Critical path of MapReduce, 256MB, \textit{burst}.}
\label{fig:eval:osnoise:map-reduce}
\end{subfigure}\hfill
\begin{subfigure}{0.322\linewidth}
\centering    
\includegraphics[width=\linewidth, keepaspectratio]{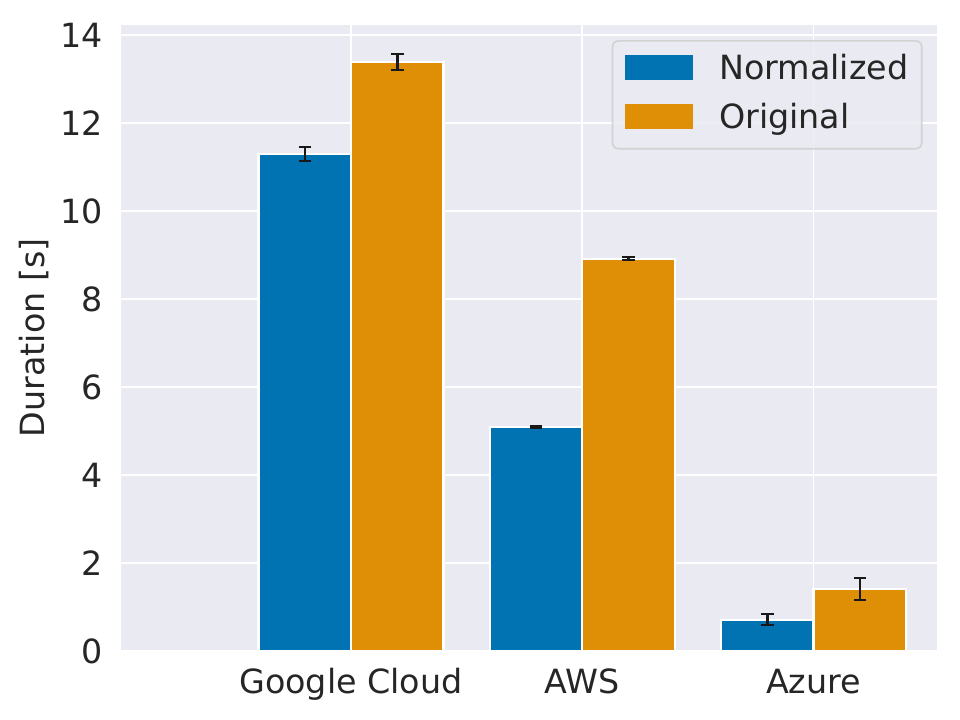}
\caption{Critical path of Machine Learning, 1024MB, \textit{burst}.}
\label{fig:eval:osnoise:ml}
\end{subfigure}
\caption{Analysis of OS noise.}
\label{fig:eval:osnoise}
\end{minipage}\hfill
\begin{minipage}{.399\textwidth}
\centering
\begin{subfigure}[b]{0.49\linewidth}
    \centering
    \includegraphics[width=\linewidth,keepaspectratio]{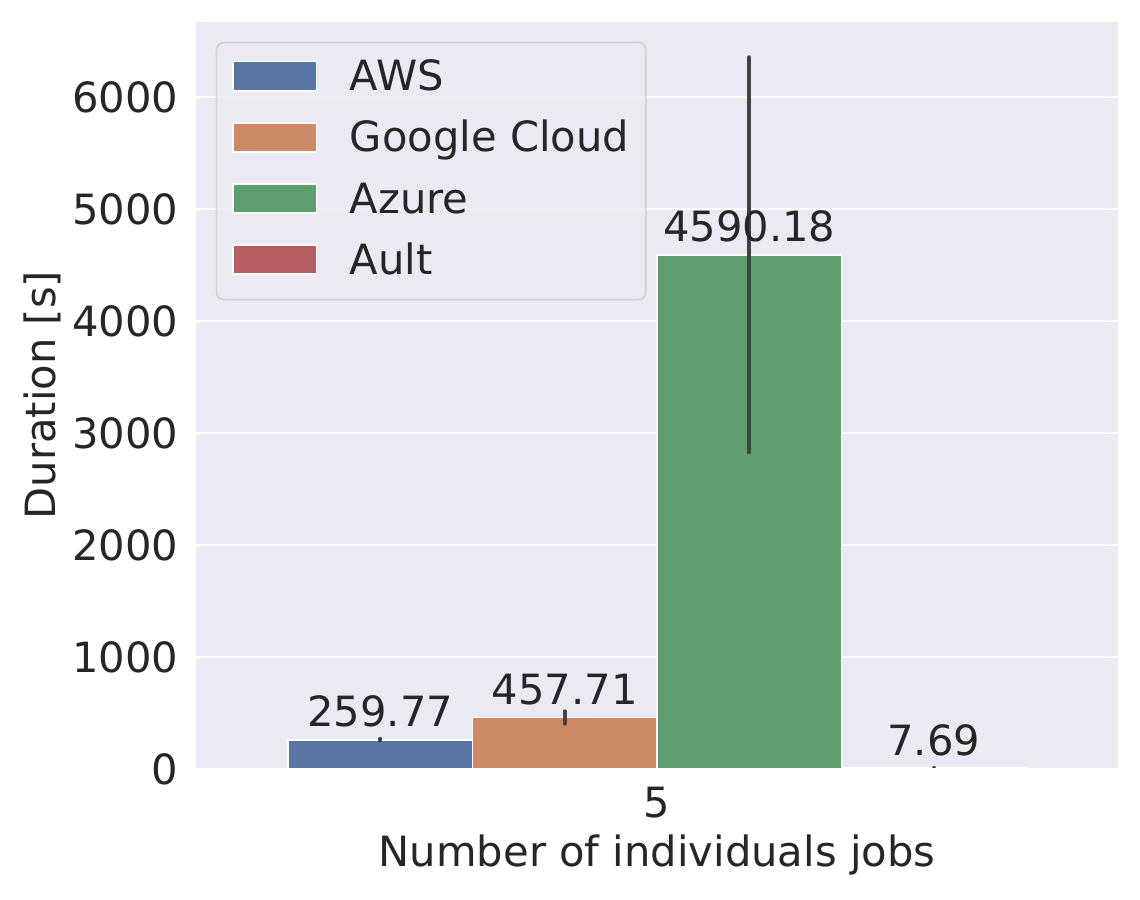}
    \caption{Complete workflow.}
    \label{fig:eval:scientific:cloud-vs-ault}
\end{subfigure}
\hfill
\begin{subfigure}{0.49\linewidth}
    \centering
    \includegraphics[width=\linewidth,keepaspectratio]{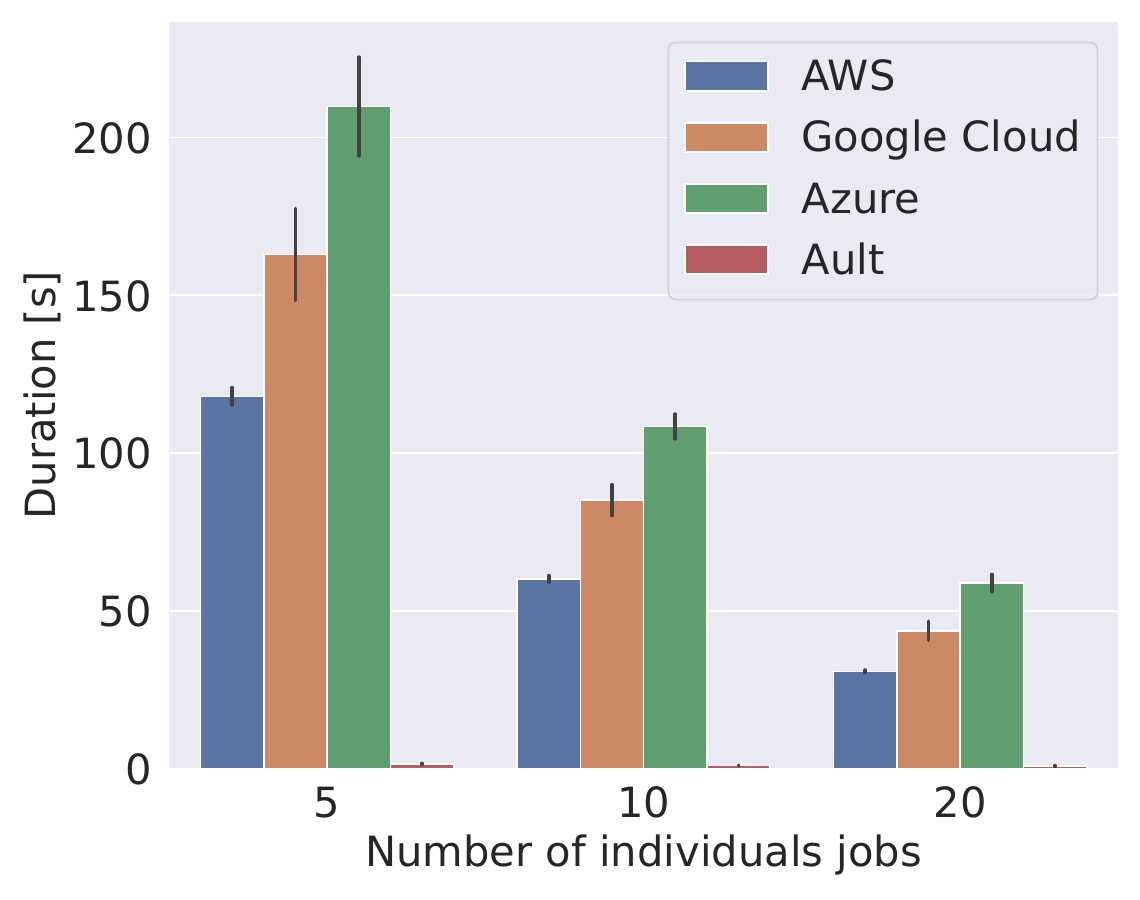}
    \caption{\texttt{individuals}.}
    \label{fig:eval:scientific:scaling-individuals}
\end{subfigure}
\caption{Scalability of 1000Genome workflow.}
\label{fig:eval:scientific}
\end{minipage}
\end{figure*}

\paragraph{OS Noise}
The cloud provider controls the CPU allocation to a serverless function, either in relation to the memory configuration on AWS and GCP~\cite{lambda-vcpu, cloud-functions-pricing}, or in an undisclosed fashion on Azure. 
We use the selfish detour benchmark to quantify OS noise~\cite{netgauge}, which allows us to estimate how long the function is suspended by the OS, which in turn approximates the vCPU timeshare.
The benchmark runs a tight loop and records the event that one iteration took significantly more cycles than expected $N$ times.
The magnitude and frequency of these events characterize the suspension and noise.
We deploy a workflow with a single function executing the benchmark, invoke it 30 times concurrently, collect $N=5000$ events, and sample warm invocations to obtain consistent results.
Figure~\ref{fig:eval:osnoise:selfish-detour} compares the relative to the expected suspension time according to the cloud documentation.
We observe less noise on Google Cloud when compared to AWS, with more than 20\% difference on 1024MB memory. 
We normalize the critical path per platform using the following approximation: given a function with memory configuration $M$, we represent the relative duration of function suspension as $S_M$ and compute the normalized critical path $T'_C = T_C * \left( 1 - S_M \right)$.
We observe the largest relative discrepancy on two benchmarks, MapReduce (Figure~\ref{fig:eval:osnoise:map-reduce}) and Machine Learning (Figure~\ref{fig:eval:osnoise:ml}). 
The overall trend observed in Section~\ref{sec:eval:performance} remains unchanged: Google Cloud demonstrates the longest critical path duration.
The suspension time explains the shorter critical path on Azure as compared to AWS and GCP for benchmarks with low-memory configurations: 
Azure functions receive larger CPU allocations.

\begin{figure}

\begin{minipage}{.49\textwidth}
\centering
\begin{subfigure}{0.49\linewidth}
    \centering
    \includegraphics[width=\textwidth, keepaspectratio]{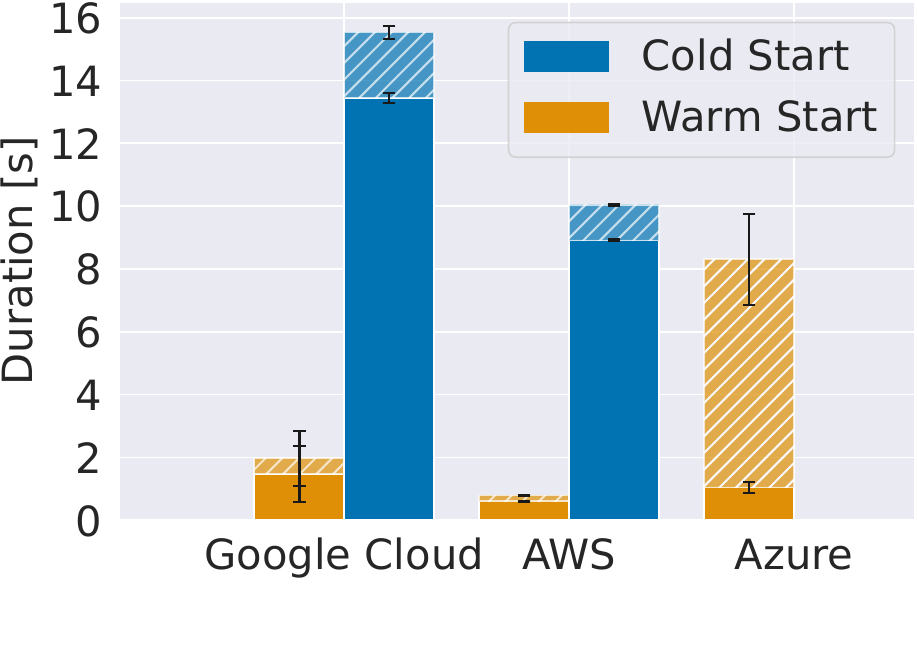}
    \caption{Machine Learning, 1024MB.}
    \label{fig:eval:cold-start:ml}
\end{subfigure}
\hfill
\begin{subfigure}{0.49\linewidth}
    \centering
    \includegraphics[width=\textwidth, keepaspectratio]{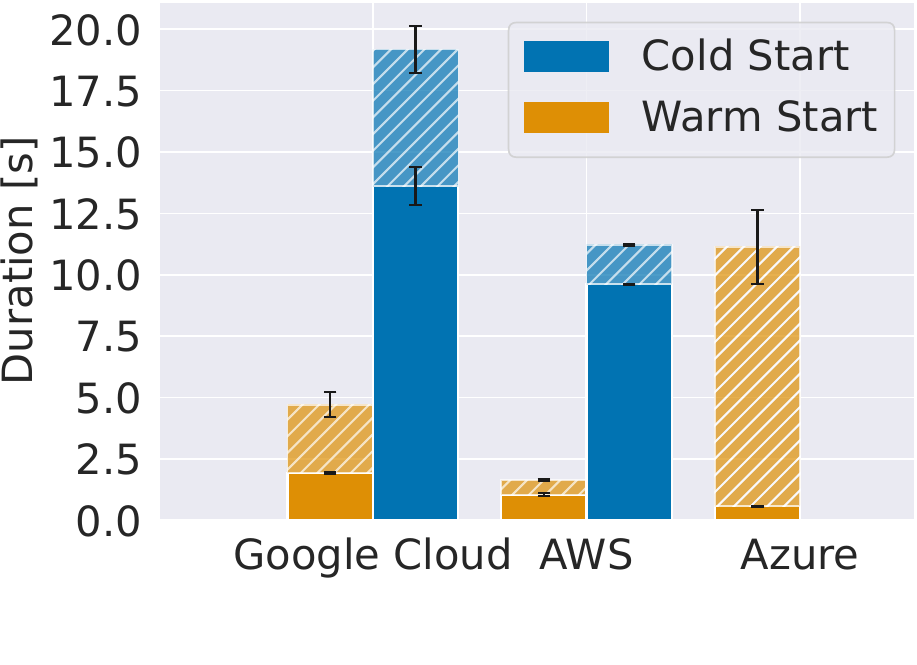}
    \caption{MapReduce, 256MB.}
    \label{fig:eval:cold-start:mapreduce}
\end{subfigure}
\caption{Critical path (opaque) and overhead (hatched) of warm and cold invocations.}
\label{fig:eval:cold-start}
\end{minipage}
\end{figure}

\paragraph{Cold Starts}
Cold invocations add significant overhead to the function execution~\cite{copik2021sebs}.
Table~\ref{tab:eval:critical:cold-starts} shows the frequency of cold starts in our measurements, 
with cold starts identified using the containerID~(see Section~\ref{sec:impl:benchmark-suite}).
Azure Durable performs significantly better, \cameraready{experiencing almost no cold starts,} likely because function apps on Azure can hold many invocations concurrently~\cite{copik2021sebs}.
While the low scalability causes high orchestration overheads, it benefits the 
computations by putting them in warm containers.
Figure~\ref{fig:eval:cold-start} shows the impact of cold starts on the critical path and overhead. 
\cameraready{We show only the duration of warm starts on Azure Durable, as all benchmarks show a very low amount of cold starts on Azure.}
\cameraready{On AWS and GCP, however, there is a high percentage of cold starts in our measurement data. We therefore} collected another 60 workflow invocations with at least one warm function and show the critical path for the resulting completely warm invocations. 
Google Cloud and AWS functions perform up to $2.0\times$ and $4.5\times$ better, respectively, achieving almost the same performance as Azure. 
Thus, cold starts are a major factor influencing the slowdown and performance instability observed in many benchmarks.

\paragraph{\eurosys{Conclusions}}
\eurosys{Our experiments show that Azure achieves short critical paths due to larger CPU allocations and less cold starts. To minimize the critical path, benchmarks using a low-memory configuration and only being executed occasionally can therefore be deployed to Azure. High-memory configurations, however, get higher CPU shares on AWS and GCP. GCP shows the slowest critical path even for warm invocations, while AWS can be competitive to Azure, making it a good choice for frequently executed workflows.}

\begin{table}
    \centering
    \small
    \begin{tabular}{l|rrr|rr}
    & \multicolumn{3}{c}{Cold starts} & \multicolumn{2}{c}{State transitions} \\
    Benchmark & AWS & GCP & Azure & AWS & GCP\\ \toprule
    Video & 86.94\% & 68.61\% & 3.89\% & 7 & 20 \\
    MapReduce & 100\% & 68.17\% & 1.0\% & 14 & 54 \\
    Trip Booking & 100\% & 38.24\% & 0.6\% & 9 & 16\\
    ExCamera & 73.58\% & 69.34\% & 0.94\% & 21 & 73 \\
    ML & 100\% & 99.26\% & 2.60\% & 6 & 18\\
    1000Genome & 98.16\% & 72.40\% & 7.72\% & 26 & 96 \\ \bottomrule
    \end{tabular}
    \caption{Relative \#cold starts and \#state transitions.}
    \label{tab:eval:critical:cold-starts}
\end{table}

\subsection{RQ3: Usability for Scientific Workflows}

There is rising interest in the scientific community to use serverless solutions~\cite{eismann2020review}, accompanied by experimentation with serverless offerings of the platforms~\cite{malawksi2020} and management systems for serverless execution of workflows~\cite{aji2019-sweep,jiang2017-dewe,10.1145/3503221.3508407,10046081}. 
However, they do not consider the workflow orchestration systems the cloud platforms offer. 
We use the scientific benchmark \textit{1000Genome} to compare cloud services and the HPC system Ault using nodes equipped with Intel(R) 6154@3.00GHz CPU, repeating measurements five times. 

First, we compare the runtime of the total workflow, as shown in Figure~\ref{fig:eval:scientific:cloud-vs-ault}.
While the workflow execution time is, on average, 457.7s and 259.8s on GCP and AWS, respectively, the execution takes only 7.7s on Ault.
GCP exhibits a coefficient of variation of 12.2\%, while AWS has a coefficient of variation of only 3.3\% - even lower than 4.1\% on Ault. 
Interestingly, I/O takes less than one second on AWS, meaning that the computation is slower in the cloud.
Then, we compare the scaling behavior of the different platforms for the \texttt{individuals} task of the workflow. We employ strong scaling, i.e., adding more jobs while keeping the size of the input file the same, resulting in smaller chunks per job. 
Figure~\ref{fig:eval:scientific:scaling-individuals} shows the speedup of 1.96 and 1.95 on AWS, 1.91 and 1.95 on GCP, and 1.51 and 1.24 on Ault for 10 and 20 jobs \cameraready{w.r.t. 5 and 10 jobs, respectively}. 
The cloud platforms achieve a nearly-optimal speedup, which is not surprising given the high overhead for the baseline execution.

\subsection{RQ4: Pricing}

\begin{figure*}[h!]
\begin{subfigure}{0.16665\linewidth}
\centering
\includegraphics[width=\linewidth]{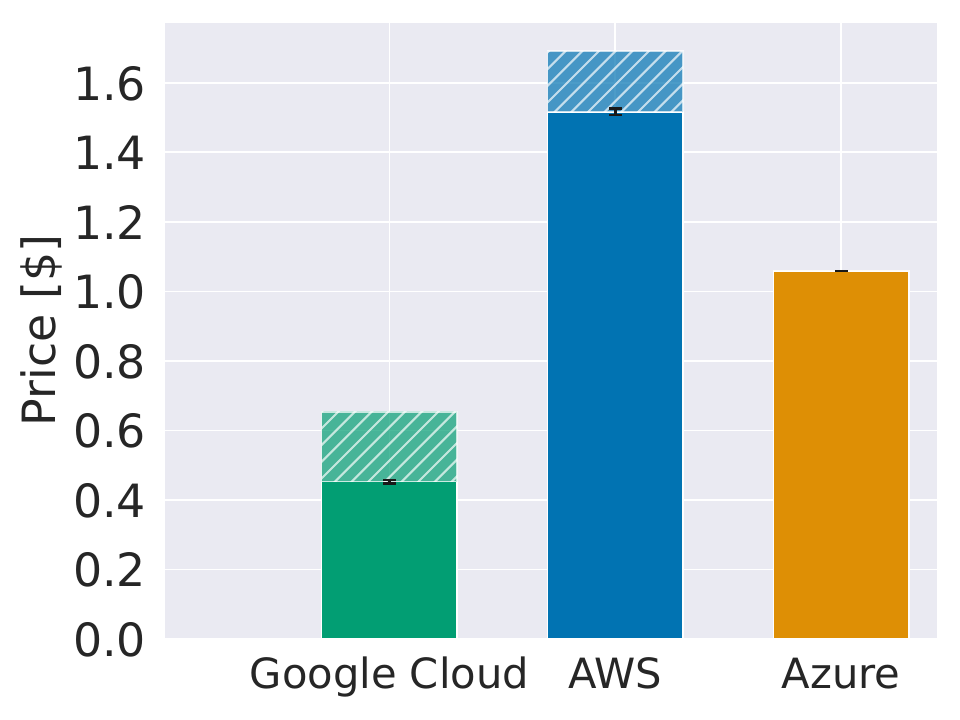}
\caption{Video Analysis}
\label{fig:eval:pricing:video-analysis}
\end{subfigure}\hfill
\begin{subfigure}{0.16665\linewidth}
\centering
\includegraphics[width=\linewidth]{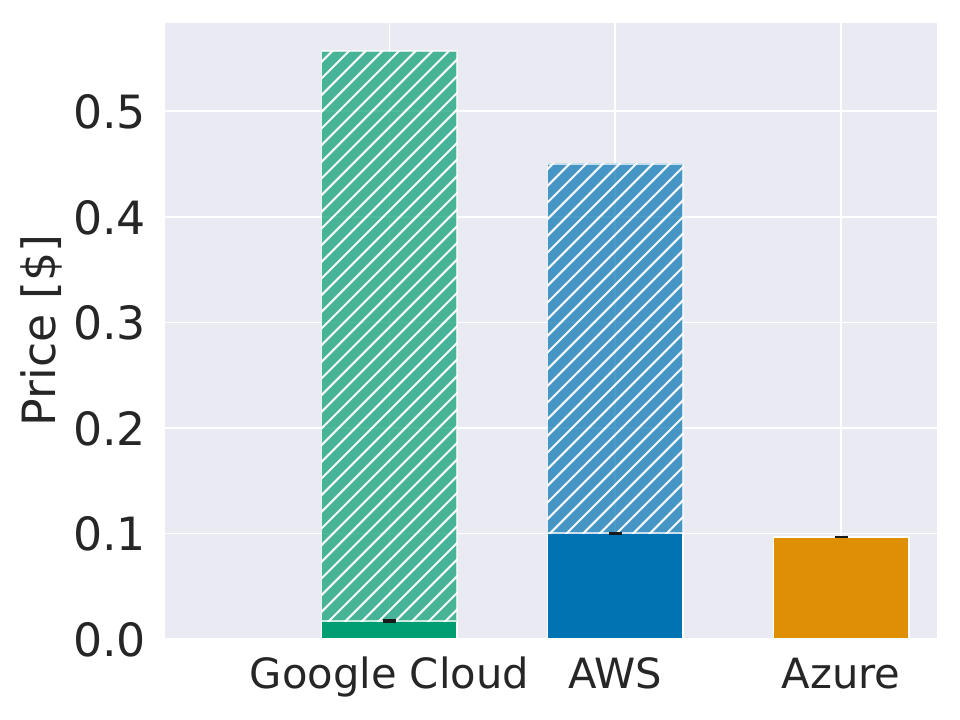}
\caption{MapReduce}
\label{fig:eval:pricing:mapreduce}
\end{subfigure}\hfill
\begin{subfigure}{0.16665\linewidth}
\centering
\includegraphics[width=\linewidth]{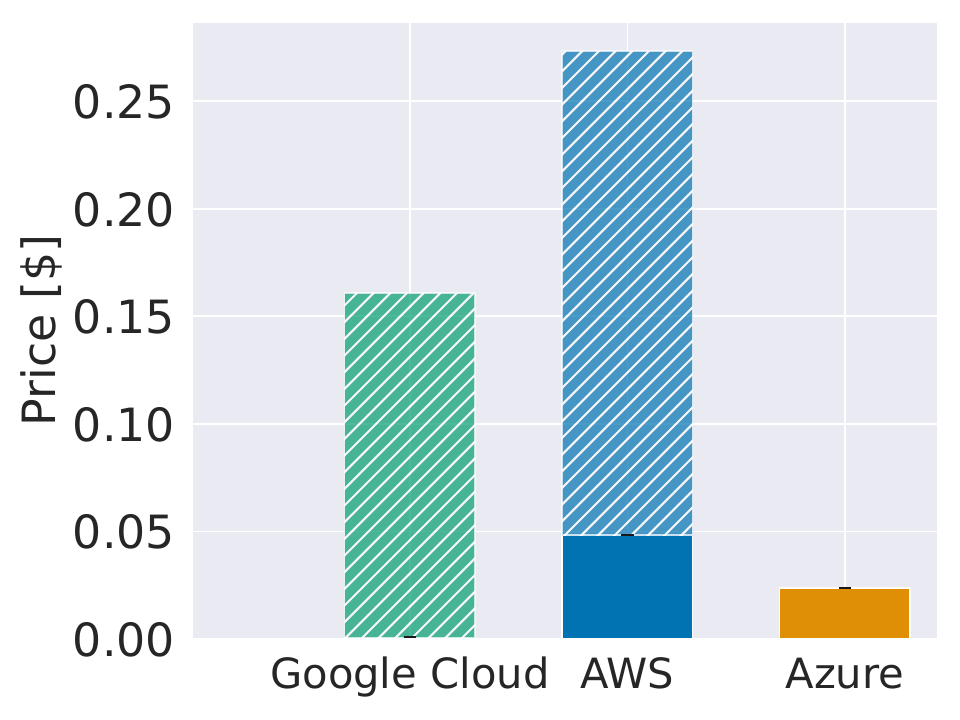}
\caption{Trip Booking}
\label{fig:eval:pricing:trip-booking}
\end{subfigure}\hfill
\begin{subfigure}{0.16665\linewidth}
\centering
\includegraphics[width=\linewidth]{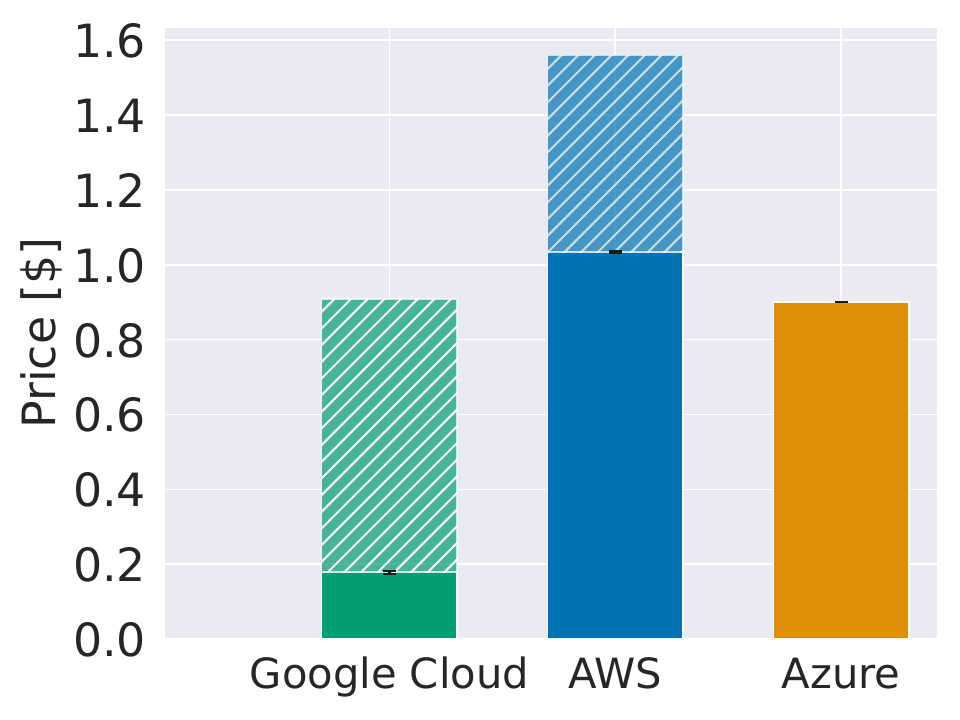}
\caption{ExCamera}
\label{fig:eval:pricing:excamera}
\end{subfigure}\hfill
\begin{subfigure}{0.16665\linewidth}
\centering
\includegraphics[width=\linewidth]{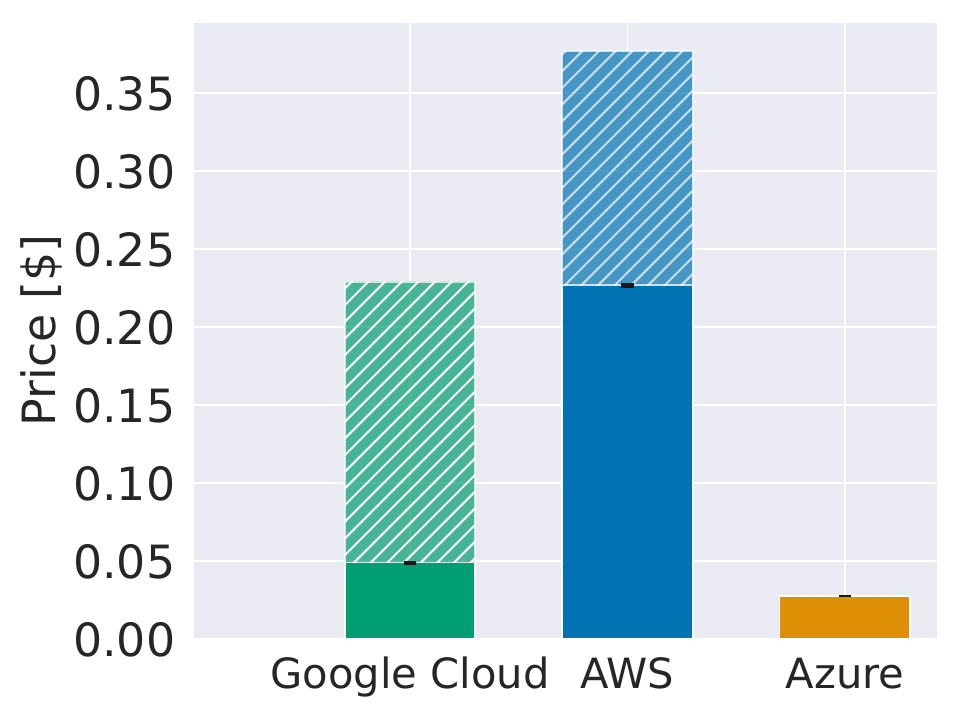}
\caption{Machine Learning}
\label{fig:eval:pricing:ml}
\end{subfigure}\hfill
\begin{subfigure}{0.16665\linewidth}
\centering
\includegraphics[width=\linewidth]{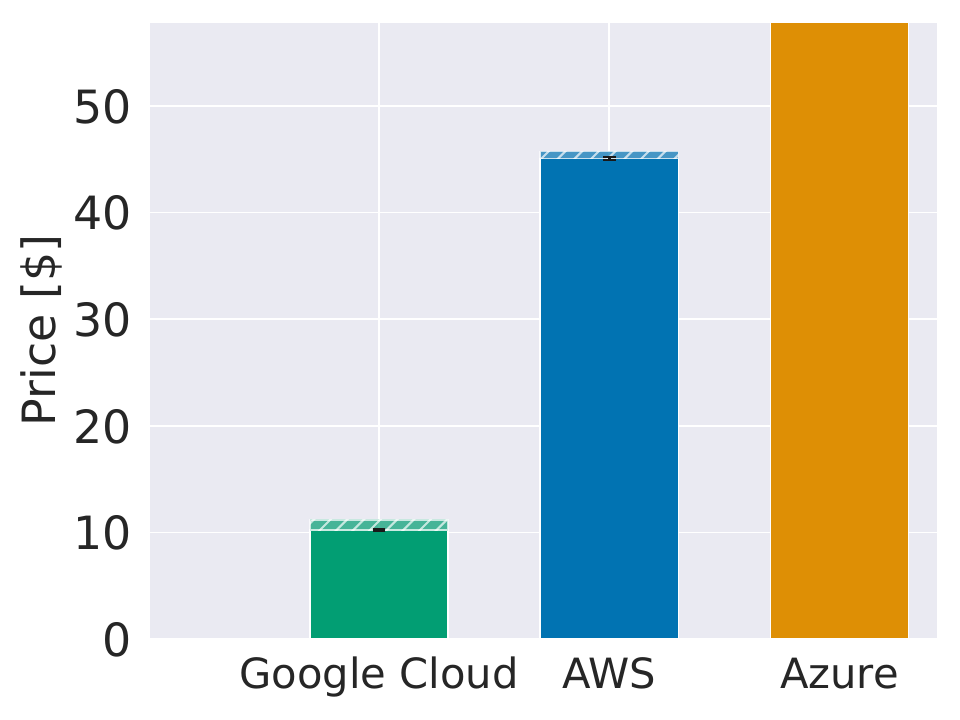}
\caption{1000Genome}
\label{fig:eval:pricing:1000genome}
\end{subfigure}
\caption{Price per 1000 workflow executions: function costs are opaque and state transition costs are translucent.}
\label{fig:eval:pricing}
\end{figure*}

We compare the average cost of executing a workflow and estimate the prices, as shown in Table~\ref{tab:platforms:pricing}, p.~\pageref{tab:platforms:pricing}.
Functions invoked during the execution of a workflow are billed based on the integral of memory and duration.
Figure~\ref{fig:eval:pricing} visualizes the cost of workflow execution split into two groups: function execution (opaque) and the cost of orchestrating the state machine (hatched). 
Note that, due to Azure's billing and measurement system, we could only retrieve an average cost value over all workflow invocations. 
Even though the Trip Booking benchmark is a simple pipeline with error catching, running it with workflow orchestration still adds significant state transition costs. 
Azure is the most expensive service for the 1000Genome benchmark. 
Google Cloud is the most expensive for MapReduce due to the high number of state transitions.
AWS Step Functions are the most expensive solution for the other four benchmarks because functions cost $6.7\times$ more for computation than Google Cloud Functions.
The price charged for state transitions is nearly identical between AWS and Google Cloud, even though AWS charges $2.5\times$ more: the AWS state language requires fewer states to implement the benchmarks (Table~\ref{tab:eval:critical:cold-starts}). 
\eurosys{Overall, we observe that Azure is expensive for workflows where it is also the slowest platform, such as for 1000Genome, but offers the cheapest pricing for benchmarks it also executes fastest, such as ML and MapReduce. Contrary to that, AWS shows high pricing for benchmarks it is fastest, such as Video Analysis and ExCamera}

In addition to execution and orchestration costs, workflows generate charges when accessing the object and NoSQL storage.
While the prices of read and write operations on the object storage are the same across clouds, the billing models for key-value storage differ: 
DynamoDB charges according to the amount of data read and written in strictly defined size increments;
CosmosDB applies the same pricing to request units but does not explicitly define expected consumption;
and Datastore has higher costs per operation but makes the cost independent of the item size.
To understand the impact of this,
we analyze the full execution of the Trip Booking benchmark.
One workflow invocation requires three insertions and three deletions, with all items taking at most a few hundred bytes.
While the estimated storage costs are similar on each platform, between \textcent 0.68 and \textcent 1.08 per 1000 executions, 
they impact the final cost differently.
NoSQL operations add only 2.74\% and 6.72\% of the total price on AWS and GCP, respectively.
The total execution cost on Azure is just \textcent 2.4.
There, the estimated cost of CosmosDB request units is equal to \textcent 0.68 and adds 28.5\% of workflow price.

\subsection{RQ5: Evolution of Performance}
\begin{figure}
\begin{subfigure}{0.4\linewidth}
\centering
\includegraphics[width=\linewidth]{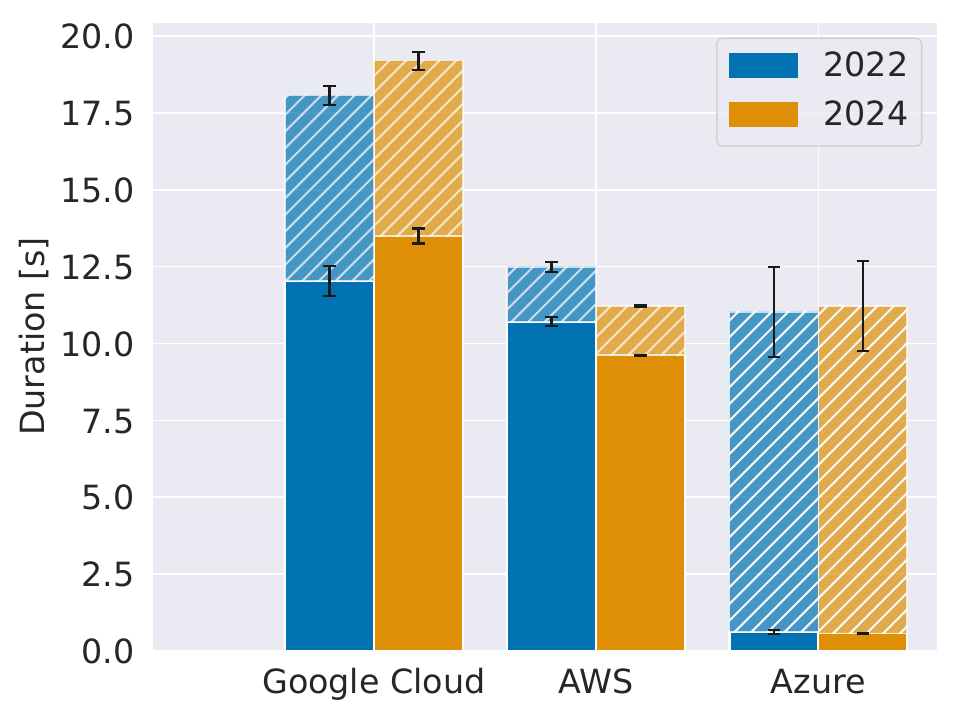}
\caption{MapReduce.}
\label{fig:eval:2022-vs-2024:mapreduce}
\end{subfigure}\hfill
\begin{subfigure}{0.4\linewidth}
\centering
\includegraphics[width=\linewidth]{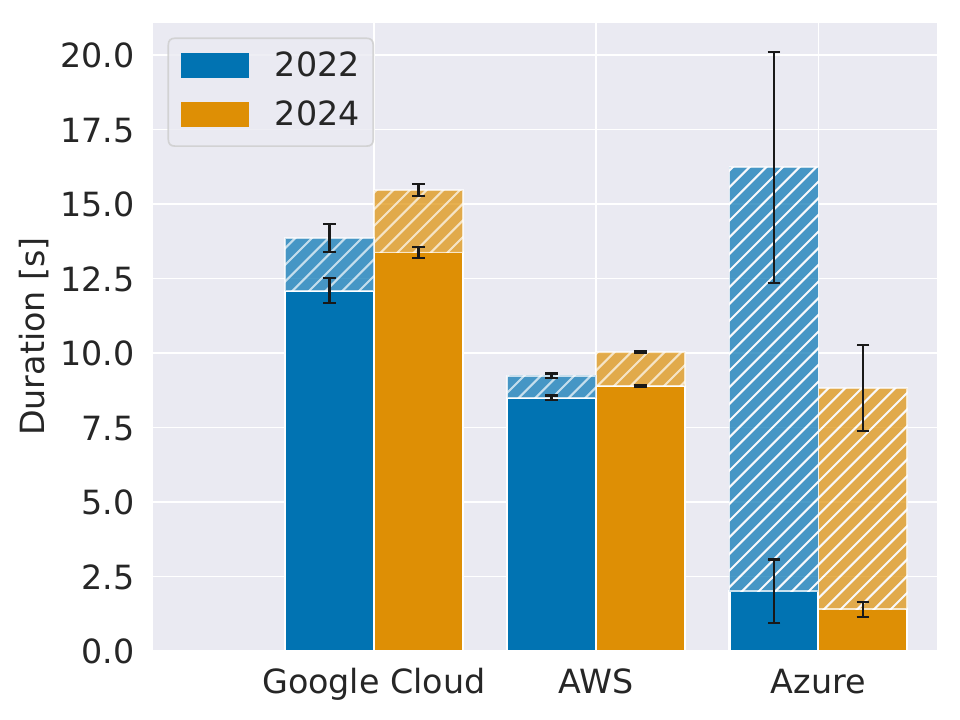}
\caption{ML.}
\label{fig:eval:2022-vs-2024:ml}
\end{subfigure}
\caption{Comparison of critical path (opaque) and overhead (hatched) between 2022 and 2024, \textit{burst} invocations.}
\label{fig:eval:2022-vs-2024}
\end{figure}

Finally, we assess the performance stability over time by comparing July 2022 and January 2024 results.
The executions from 2022 contain 30 invocations per workflow using Python 3.7, in cloud regions \emph{europe-west} for Azure, \emph{europe-west-1} for GCP, and \emph{us-east-1} for AWS.
The 2024 invocations are run in the same regions, except for GCP in \emph{us-east1}, and use Python 3.8.
Figure~\ref{fig:eval:2022-vs-2024} shows the results. 
The critical path and overhead of the MapReduce and ML benchmark are approximately the same on Google Cloud. 
The runtime on AWS is quite stable without any notable differences between 2022 and 2024.
Azure has a stable duration of the critical path.
While the overhead for MapReduce is the same in 2024 as in 2022, the overhead of ML has been approximately halved from 2022 to 2024.

\subsection{Threats to Validity}

A threat to the external validity is our choice of benchmark applications. We mitigate this by using applications from different domains that correspond to previous findings on the characterization of workflow use cases~\cite{eismann2020review,eismann2021review}.
Regarding internal validity, the different geographical regions and different week days we conducted our measurements on could have an impact. 
While we repeat each experiment six times to obtain stable results, there could be performance variability based on the time of day. 
However, systematically investigating this is beyond the scope of our work. 
\section{Related Work}

Multiple benchmark suites have been proposed to cover different aspects of serverless computing, from microarchitecture to the application level~\cite{tail-latency,faas_arch,copik2021sebs,faasbenchmark,Back2018,Kim2019,10.1145/3401025.3401738}. \changed{However, all of them consider only the execution of single functions.}
\changed{\citet{Das2018} benchmark serverless edge computing platforms. 
Other performance studies of serverless applications focus on non-workflow orchestration systems, e.g., using cloud storage and queue triggers~\cite{scheuner2022lets,9610428,9860528,9027346}. 
\citet{9610428} propose BeFaaS, providing an application benchmark modeling an online shop where the functions communicate using synchronous and asynchronous calls. 
In contrast, \toolname{} targets serverless workflow orchestrations.} 

\eurosys{
ServerlessBench~\cite{10.1145/3419111.3421280} considers a function chain microbenchmark orchestrated by AWS Step Functions, but only measures runtime of the workflow and time in between function invocations for varying payload sizes. 
Kousiouris et al.~\cite{10.1145/3491204.3527467} use microbenchmarks to estimate the overhead of orchestration in OpenWhisk.
López et al.~\cite{8605772} investigate the orchestration overhead with microbenchmarks of function chains and parallel functions.
Shahidi et al.~\cite{9668304} evaluate the performance and cost of two stateful workflows on AWS and Azure.
Barcelona-Pons et al.~\cite{10.1145/3366623.3368137} use a microbenchmark to test the performance of fork-join parallelism in workflow orchestrators.
With \toolname{}, we provide not only microbenchmarks, but also six applications from different domains that can automatically be deployed to different cloud platforms. Based on these benchmarks, we present a broader evaluation of the performance of cloud platforms.
}

Wen et al.~\cite{serverless-study} conducts a performance investigation of serverless workflows using two applications and microbenchmarks with varying numbers of functions, payload size, and parallelism.
While they measure the execution time and estimate overhead, they do not evaluate scalability, billing, or investigate overhead sources.
Instead, we focus on a wider collection of applications and propose a unifying model that allows developers to deploy and evaluate a single implementation across many cloud platforms. 
Moreover, we make all benchmark codes available and provide a ready-to-use benchmarking platform.
Finally, we evaluated serverless Google Cloud Workflows instead of the non-serverless Google Cloud Composer.
\eurosys{XFBench~\cite{xfbench} provides chaining of different functions and deploying them to AWS Step Functions and Azure Durable Functions, while we focus on realistic and complete applications. 
Moreover, they do not consider cloud-native data movement between functions via cloud storage, do not evaluate the overhead of their platform transcription, and can not compare pricing between platforms.}

Other authors analyzed the productivity of workflow languages and proposed alternative models.
AFCL~\cite{afcl} is a custom and provider-independent orchestration language for serverless workflows, implemented on top of AWS Step Functions and IBM Composer.
Burckhardt et al. explore the semantics of Durable Functions~\cite{durable-functions-semantics} and propose Netherite~\cite{netherite}, a new engine to replace Azure Durable Functions.

\section{Conclusions}
We propose \toolname{}, the first benchmark suite for serverless workflows.
We follow the established benchmark design principles: introduce a platform-agnostic workflow model, propose a collection of six representative applications, and integrate them into an existing benchmark suite to ensure reproducibility and ease of use.
We support the three major cloud providers, and benchmarks can be ported to other services by implementing a single interface transcribing our model to the cloud-specific interface.
We conduct a comprehensive and long-term evaluation of the performance and cost of proposed benchmark applications,
investigating factors influencing the runtime and variance: cold startups, 
noise, scheduling, and the storage I/O.
With the new benchmark suite, we enable benchmarking of the same workflow on different platforms, \changed{providing software developers and researches with valuable insights regarding their different behaviors and properties.} 

\ifARXIV
\section*{Acknowledgments}
\setlength{\intextsep}{1pt}
\setlength{\columnsep}{5pt}
\begin{wrapfigure}{r}{.13\linewidth} 
  \includegraphics[width=.13\columnwidth]{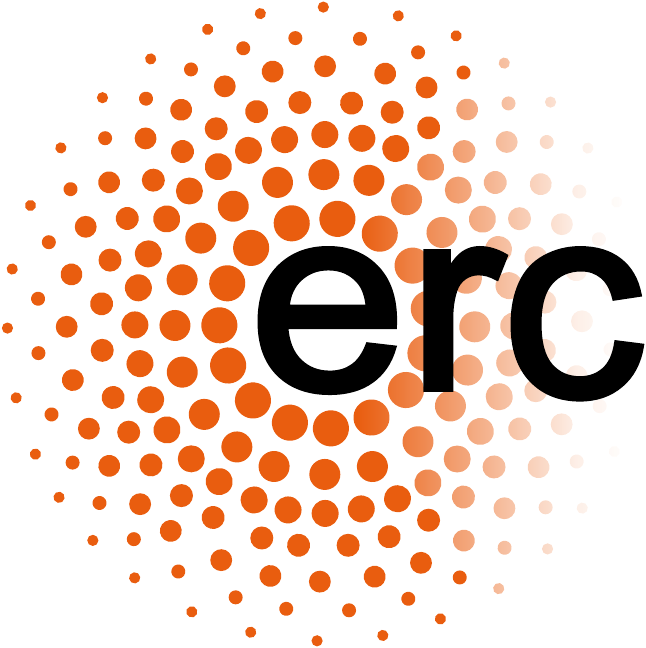}
\end{wrapfigure}
Larissa Schmid is supported by the pilot program Core Informatics at KIT (KiKIT) of the Helmholtz Association (HGF).
This project has received funding from the European Research Council (ERC) under the European Union’s Horizon 2020 program (grant agreement PSAP, No. 101002047).  We would also like to thank the Swiss National Supercomputing Centre (CSCS) for providing us with access to their HPC machine Ault. We thank Amazon Web Services for supporting this research with credits through the AWS Cloud Credit for Research, and Google Cloud Platform through the Google Cloud Research Credits program with the award GCP19980904.
\fi

\section*{Authorship Statement}
\noindent 
The authors contributed to the paper as follows: M. Copik,  A. Calotoiu, and T. Hoefler conceived the initial idea and L. Schmid, M. Copik,  A. Calotoiu, and T. Hoefler designed the study; L. Brandner implemented the initial model, and L. Schmid extended and formalized it; L. Brandner implemented the benchmarks and L. Schmid and M. Copik extended and improved the implementation; L. Schmid and M. Copik collected data;  L. Schmid, M. Copik, and L. Brandner analyzed and interpreted the results; L. Schmid and M. Copik conducted the literature study; L. Schmid and M. Copik wrote the draft manuscript; and L. Schmid, M. Copik, A. Calotoiu, A. Koziolek, and T. Hoefler reviewed and revised the manuscript.

\bibliographystyle{ACM-Reference-Format}
\bibliography{refs,serverless.merged,cloud.merged,hpc.merged,own}

%
\clearpage
\newpage

\appendix
\section{Artifact Appendix}

\subsection{Abstract}

Our artifact contains the implementation of \toolname{}, data, and analysis scripts. 
We provide the following components:
\begin{itemize}
    \item \textit{sebs-flow-implementation} - Source code of the benchmark suite.
    \item \textit{sebs-flow-artifact} - Benchmarking results obtained for the paper together with Python plotting and analysis scripts used for data analysis.
\end{itemize}

\subsection{Description \& Requirements}


\subsubsection{How to access}
\hfill \break
\url{https://doi.org/10.5281/zenodo.14809924}

\subsubsection{Hardware dependencies}

Unix system capable of sending HTTP requests to the cloud.

\subsubsection{Software dependencies}

\begin{itemize}
    \item Docker (at least 19)
    \item Python 3.7+ with \texttt{pip} and \texttt{venv}
    \item \texttt{libcurl} and its headers must be available to install \texttt{pycurl}
    \item Standard Linux tools and \texttt{zip} installed
\end{itemize}

\subsubsection{Benchmarks} 

We provide benchmarks and data used as part of the implementation in \texttt{server\-less-\-bench\-marks/\-benchmarks/\-600.workflows} and \texttt{ser\-ver\-less-\-bench\-marks/\-benchmarks-\-data/\-600.workflows}. 

\subsection{Set-up}

To install the benchmark suite with support for all platforms used for our evaluation, use \texttt{./install.py --aws --azure --gcp}. This will create a virtual environment in \texttt{python-venv}, and install necessary Python dependencies and third-party dependencies. To use \toolname{}, the new Python virtual environment has to be activated: \texttt{. python-venv/bin/activa\-te}. 
To deploy benchmarks to a platform, account credentials musst be supplied. See \texttt{serverless-benchmarks/\-docs/\-plat\-forms.md} for details. 

For measuring the execution of serverless workflows, we use a \textit{Redis} instance deployed on a VM in the same cloud region as the workflow and its resources. See \texttt{serverless\-bench\-marks/\-docs/workflows.md} for details. 

\subsection{Evaluation workflow}

\subsubsection{Major Claims}

\noindent
\textit{RQ1: Runtime.} We evaluate the runtime differences between platforms by executing Experiment E1. Results are shown in Figure~\ref{fig:eval:performance} and~\ref{fig:eval:overhead} and discussed in Section~\ref{sec:eval:performance}. 

\noindent
\textit{RQ2.1: Orchestration Overhead.} We evaluate the overheads caused by cloud storage I/O via executing Experiment E3 (Figure~\ref{fig:eval:overhead}), by parallel scheduling via executing Experiment E4 (Figure~\ref{fig:eval:overhead:parallelsleep}), and by the return payload via Experiment E5 (Figure~\ref{fig:eval:overhead:function-chain}). 

\noindent
\textit{RQ2.2: Critical Path Discrepancy.} We evaluate how the critical path is impacted by OS noise via executing Experiment E6 (Figure~\ref{fig:eval:osnoise}) and using data from E1 (Figure~\ref{fig:eval:osnoise:map-reduce},~\ref{fig:eval:osnoise:ml}) and by cold starts by analyzing the results from E1. 

\noindent
\textit{RQ3: Usability for Scientific Workflows.} We evaluate how well serverless workflow orchestrations are suited for execution of scientific benchmarks by comparing execution times and scaling of the workflow on cloud platforms and an HPC system using data from E1 for 1000Genomes and with Experiments E7 and E8 (Figure~\ref{fig:eval:scientific}). 

\noindent
\textit{RQ4: Pricing.} We evaluate the differences in pricing between the cloud platforms by comparing execution cost of our application benchmarks using data from E1 (Figure~\ref{fig:eval:pricing}). 

\subsubsection{Experiments}

All configuration files used are provided per platform and benchmark executed as part of the artifact. 


\noindent
\textit{E1: Burst execution of application benchmarks.} Execution of all application benchmarks. The paper uses 180 \textit{burst} workflow executions per benchmark with 30 executions triggered at once. 

\noindent
\textit{E2: Warm execution of application benchmarks.} Execution of Machine Learning and MapReduce benchmarks in \textit{warm} mode. The paper collects 60 workflow executions with at least one \textit{warm} function invocation per benchmark with 30 executions triggered at once. 

\noindent
\textit{E3: Parallel Download.} Execution of the parallel download microbenchmark with 20 functions downloading a file in parallel, with filesizes from $2^{10}b$ to $2^{28}b$. The paper uses 30 \textit{burst} executions triggered at once. 

\noindent
\textit{E4: Parallel Sleep.} Execution of the parallel sleep microbenchmark with $2 \leq N \leq 16$ functions and sleep durations of $1 \leq T \leq 20$s. The paper uses 30 \textit{burst} executions triggered at once. 

\noindent
\textit{E5: Function Chain.} Execution of the function chain micro\-bench\-mark with 10 functions, with functions returning $2^{5} \leq M 2^{18}b$ sent to the next function. The paper uses 30 \textit{warm} executions. 

\noindent
\textit{E6: OS Noise.} Execution of the selfish detour microbenchmark collecting $N = 5000$ events, executed with memory configurations from 128MB to 2048MB. The paper uses 30 \textit{warm} executions.

\noindent
\textit{E7: Execution of 1000Genomes on HPC system.} Execution of the 1000Genomes workflow on an HPC system. The paper uses five repetitions per configuration and a Intel(R) 6154@3.00GHz CPU. To reproduce the result that the execution on an HPC system is much faster, however, the workflow can be run in virtually every HPC environment. 

\noindent
\textit{E8: Scaling of \texttt{individuals} task of 1000Genomes workflow.} Execution of the workflow \texttt{6101.\-1000-\-ge\-nome-\-indi\-vi\-du\-als} with inputs \textit{small-10} and \textit{small-20}. The paper uses 180 \textit{burst} workflow executions per input with 30 executions triggered at once. 

\subsection{Notes on Reusability}
\label{sec:reuse}

\toolname{} can be extended to evaluate workflow orchestrations on new serverless platforms. Moreover, new workflow benchmarks can be added to evaluate their performance on different platforms. Repetition of benchmarks over time can give insights into the evolution of performance. 




\end{document}